\documentclass[manuscript]{acmart} 

\AtBeginDocument{%
  }

\setcopyright{acmlicensed}
\copyrightyear{2018}
\acmYear{2018}
\acmDOI{XXXXXXX.XXXXXXX}
\acmConference[Conference acronym 'XX]{Make sure to enter the correct
  conference title from your rights confirmation email}{June 03--05,
  2018}{Woodstock, NY}
\acmISBN{978-1-4503-XXXX-X/2018/06}

\usepackage{xcolor}
\usepackage{tabularx}
\usepackage{placeins}
\usepackage{multirow}
\usepackage{colortbl}
\usepackage{svg}
\usepackage[most]{tcolorbox}

\newcommand{\eg}[0]{{e.g.,~}}
\newcommand{\ie}[0]{{i.e.,~}}

\newcommand{\vs}[0]{vs.}
\newcommand{\fig}[1]{Figure~\ref{#1}}
\newcommand{\tab}[1]{Table~\ref{#1}}
\newcommand{\sect}[1]{Section~\ref{#1}}
\newcommand{\appref}[1]{Appendix~\ref{#1}}

\newcommand{\qt}[1]{\textit{``#1''}}

\definecolor{ysspurple}{RGB}{102,0,102}
\definecolor{yssdgreen}{RGB}{0,100,0}
\definecolor{yssteal}{RGB}{0,128,128}
\definecolor{yssnavy}{RGB}{0,0,128}
\definecolor{ysspink}{RGB}{255,20,147}
\definecolor{yssgreen}{RGB}{107,142,35}
\definecolor{yssgray}{RGB}{47,79,79}
\definecolor{yssbluepurple}{RGB}{138,43,230}
\definecolor{yssolive}{rgb}{0.33, 0.42, 0.18}

\definecolor{rowgray}{gray}{0.82}

\newif{\ifhidecomments}
\hidecommentstrue
\ifhidecomments
    \newcommand{\ycomment}[1]{}
    \newcommand{\yasaman}[1]{}
    \newcommand{\fatemeh}[1]{}
\else
    \newcommand\ycomment[1]{\textcolor{ysspurple}{\textit{Yasaman: #1}}}
    \newcommand\yasaman[1]{\textcolor{ysspurple}{\textit{#1}}}
    \newcommand\fatemeh[1]{\textcolor{yssnavy}{\textit{[FM: #1]}}}
\fi

\newcommand\wb{well-being}

\newcommand\selftracking{self-tracking}
\newcommand\taskex{task examples} 

\newcommand\ataskex{a task example} 
\newcommand\onetaskex{task example} 
\newcommand\tasks{tasks} 
\newcommand\Tasks{Tasks} 
\newcommand\Task{Task}
\newcommand\task{task}
\newcommand\TaskEg{Task Example}
\newcommand\Taskeg{Task example}
\newcommand\taskeg{task example}
\newcommand\evalsetting{evaluation setting}
\newcommand\Evalsetting{Evaluation setting}
\newcommand\EvalSetting{Evaluation Setting}
\newcommand\settingsDIM{dimension}
\newcommand\phinfo{personal health informatics}
\newcommand\sut{system under test}

\newcommand{\healthbench}{HealthBench}
\newcommand{\phia}{PHIA}
\newcommand{\lifeagentbench}{LifeAgentBench}
\newcommand{\dmbench}{DM-Bench}
\newcommand{\cgmspeak}{CGM Speak}
\newcommand{\perqa}{PerQA}
\newcommand{\pha}{PHA}
\newcommand{\llmcgm}{LLM-CGM}

\newcommand{\nBenchmarks}{eight} 
\newcommand{\nCorpusBehaviorVar}{seven} 
\newcommand{\nPublishedBenchmarkTasks}{53,587} 
\newcommand{\nPHIBenchmarkTasks}{27,801} 
\newcommand{\nBehaviorVar}{86.6\%}

\newcommand{\nDataQual}{0.2\%}
\newcommand{\nPrep}{\textless{}0.1\%}
\newcommand{\nSnowball}{268} 
\newcommand{\nAIlabeledIncluded}{60} 
\newcommand{\searchEndDate}{June 2026} 

\newcommand{\taskBehaviorVar}{Behavior Variations}
\newcommand{\taskBehaviorvar}{Behavior variations}
\newcommand{\taskbehaviorvar}{behavior variations}
\newcommand{\taskCorrelation}{Correlation Analysis}

\newcommand{\taskcorrelation}{correlation analysis}
\newcommand{\taskComparison}{Comparison}
\newcommand{\taskcomparison}{comparison}
\newcommand{\taskDataQual}{Data Quality Assessment}

\newcommand{\taskdataqual}{data quality assessment}
\newcommand{\taskPItoAdvice}{Advice Grounded in Personal Data}

\newcommand{\taskpitoadvice}{advice grounded in personal data}
\newcommand{\taskPrep}{Preparation and Setup}

\newcommand{\taskprep}{preparation and setup}

\begin{document}

\title{How Well Do Benchmarks Evaluate LLMs' Use for Personal Health Informatics?}

\author{Fatemeh Mirhosseini}
\email{fatemeh.mirhosseini@tamu.edu}
\orcid{0009-0009-3593-726X}
\affiliation{%
  \institution{Texas A\&M University}
  \city{College Station}
  \state{Texas}
  \country{USA}}

\author{Yasaman S. Sefidgar}
\email{yasaman@tamu.edu}
\orcid{0000-0001-8990-699X}
\affiliation{%
  \institution{Texas A\&M University}
  \city{College Station}
  \state{Texas}
  \country{USA}}

\renewcommand{\shortauthors}{Mirhosseini et al.}

\begin{abstract}
Benchmarking is a valuable technique for advancing the capabilities of Large Language Models~(LLMs), including in personal health informatics.
There are growing efforts on benchmark creation for personal health data activities.
However, it is unclear whether and how well these efforts align with needs and considerations of personal health informatics process.
Addressing this need, we first analyzed \nPublishedBenchmarkTasks\ tasks across \nBenchmarks\ benchmarks related to personal health informatics and characterized the space of benchmark design.
We then contrasted our observations of design choices with personal informatics theories and empirical knowledge.
We find significant mismatches between existing benchmark design and domain guidelines and understanding.
This lack of alignment limits the utility of the evidence current benchmarks offer about LLM capabilities.
We use domain knowledge to highlight capabilities benchmarks should include and discuss opportunities our work enables for taking a human-centered and domain-driven approach to benchmark design. 
\end{abstract}




\received{1 November 2026}
\received[revised]{xx Month yyyy}
\received[accepted]{xx Month yyyy}

\maketitle

\section{Introduction} 
\label{sec:intro}
Recent advances in Artificial Intelligence (AI) and the advent of Large Language Models (LLMs) can address long standing challenges in personal health informatics. There is promising evidence that LLMs can support individuals in deciding what data to collect~(\eg \cite{Ishita:2026Casebot}), how to collect~(\eg \cite{Kim:2024Mindfuldiary, Hoefer:2025Telltime, Chen:2026Rhyming}) 
and make sense of it~(\eg \cite{Stromel:2024Narrating, Loerakker:2025Giveandtake}) and, when relevant, act on data-driven insights~\cite{Wu:2024Mindshift, Jorke:2025Gptcoach, Jorke:2026Bloom, Wang:2025Sleep}. 
However, much of the evidence comes from studies examining a small number of LLM capabilities, for a limited scope of tasks, settings, and contexts, often within a convenience sample. Neither use~cases nor the samples are representative of the typical diversity present in personal health informatics. Moreover, these reports are hard to compare and synthesize across studies and rarely consider LLM capabilities holistically. 
While driving discovery and innovation, this approach to evaluating LLMs leaves much to desire in understanding and establishing their capabilities.

LLM capabilities can also be examined through benchmarks. Benchmarking can complement small-scale evaluations by expanding the coverage of capabilities, tasks, settings, and contexts often in considerably larger samples. Benchmarks can provide a consistent set of tests, protocols, and metrics for standardized evaluations that profile LLM system capabilities and enable comparative analysis of different systems. These qualities can drive progress and have already prompted the creation of benchmarks for personal health informatics, \eg for diabetes management ~\cite{Cardei:2025Dmbench, Healey:2024LLMcgm}, fitness and sleep wearable data analysis ~\cite{Merrill:2026Wearable}, and personalized health recommendation ~\cite{Heydari:2025Anatomy}.

However, the reliability of benchmark evidence on LLM capabilities heavily depends on specifics of benchmark design~\cite{Liu:2024Ecbd}. Benchmarks may be detached from the reality of the activities they target~\cite{Wang:2026Agentwork} or fail to make proper measurements~\cite{Liu:2024Ecbd}. As benchmarking is increasingly considered in generating evidence for LLM use for personal health informatics, it is important to assess benchmarks supporting personal informatics activities, document their shortcomings, and pursue principled approaches to benchmark design suited to the specifics of the domain. 

In this paper, we investigate whether and how well existing benchmarks establish LLM capabilities for personal health informatics activities. We examine \nPublishedBenchmarkTasks\ tasks across \nBenchmarks\ benchmarks related to personal informatics, map out existing benchmark design and characterize the space of design choices. 
Comparing and contrasting existing benchmarks with the rich literature on personal informatics through this characterization, we identify gaps and opportunities in benchmark design that can be systematically addressed by leveraging theories and knowledge of personal informatics. 
We make the following contributions:

\begin{itemize}
    \item We characterize the space of benchmark design for personal informatics, offering a resource that can be used to create new benchmarks, analyze existing ones, and define boundaries of their applicability. 
    \item We identify significant gaps in current state of benchmark design that limit their ability to demonstrate LLM system capabilities for supporting personal health informatics activities in real-world contexts.
    \item We discuss how domain knowledge and human-centered techniques can guide benchmark design choices, from the choice of capabilities to the design of tasks and evaluation settings that demonstrate those capabilities. 
\end{itemize}

We review personal informatics process as well as needs and considerations for benchmark design and assessment in this domain in \sect{sec:back}, detail our benchmark selection and analysis process in \sect{sec:method}, and share findings on the current landscape of personal informatics benchmarking and notable gaps 
in \sect{sec:results}. We conclude by discussing how our findings can be leveraged for human-centered benchmark development.

\section{Background}
\label{sec:back}
We review the process of personal informatics~(\sect{sec:back-model}) and evidence for LLMs' support of the process~(\sect{sec:back-llms}) to inspire the range of evaluation needs to consider in benchmarking personal health informatics. We also position our work in the broader space of principled approaches to benchmark design~(\sect{sec:back-benchmarks}).

\subsection{Models of Personal Informatics}
\label{sec:back-model}
Models of personal informatics describe the process by which individuals record personally relevant information to support various needs including understanding and management of their health and \wb. 
\citet{Li:2010Stage} identified five distinct but inter-related stages of personal informatics: (1)~preparation when people decide what to track and how, (2)~collection when people use various tools and record data, (3)~integration when they gather different sources of data and analyze them for the next stage, (4)~reflection when people review their data and gain insights, and (5)~action when they leverage the new found insight toward their needs. 
\citet{Epstein:2015Lived} extended this model to also include lapsing and resumption to better reflect the lived experience of personal informatics. The stages are inter-related in that barriers in earlier stages cascade and impact later stages. Moreover, people often go back and forth across the stages. 
Furthering the representation of the lived experience in personal informatics, \citet{Sefidgar:2024Migrainetracker} connected individuals' goals to the process. They demonstrated that individuals often simultaneously pursue multiple evolving goals and successful tracking depends on alignment and realignment of different classes of goals with each other and with other aspects of tracking including \textit{tracking models}~(\ie when recording happens) and \textit{data models}~(\ie the unit of recording). 
Overall, these models distill information about \textit{what} happens in personal tracking and \textit{how}. As such, they point out the capabilities LLMs should offer and thus be evaluated for. 
We draw on this knowledge to investigate if existing benchmarks can adequately evaluate LLMs for personal health informatics activities.



\subsection{LLMs for Personal Health Informatics}
\label{sec:back-llms}
With their flexibility in conversational interaction, LLMs have shown promise in supporting \phinfo\ across the stages of \selftracking. 
Recent work has explored a range of LLM capabilities including the ability to 
help individuals plan \selftracking\ for self-experimentation~\cite{Ishita:2026Casebot}, 
more easily collect self-reported data~\cite{Chen:2026Rhyming, Hoefer:2025Telltime, Kim:2024Mindfuldiary}, 
analyze and make sense of wearable data~\cite{Stromel:2024Narrating, Loerakker:2025Giveandtake} at opportune times~\cite{Nepal:2024Mindscape},
and plan actions for regulating physical activity~\cite{Jorke:2025Gptcoach, Jorke:2026Bloom}, sleep~\cite{Wang:2025Sleep}, and problematic smartphone use~\cite{Wu:2024Mindshift}.
%

However, these research studies primarily performed focused assessments of their proposed features and interactions at a scale suited to their purpose, rather than establishing broad generalization across settings.
Establishing these capabilities for the diverse range of needs in personal health informatics thus calls for scalable evaluation methodologies. 
Moreover, LLMs change rapidly and should be continuously re-evaluated to keep our understanding of their capabilities and behaviors up to date. 
Evaluations must also go beyond purpose-built systems of these studies, as many people turn to general-purpose LLMs for health questions and increasingly connect them to their personal health data. 
Evaluation via benchmarking can address these needs 
and offer complementary evidence for LLM systems' ability to demonstrate capabilities of interest.  
In this work, we examine to what extent benchmarking extends the preliminary evidence of small-scale studies across systems and contexts.

\subsection{Design of LLM Benchmarks}
\label{sec:back-benchmarks}

Benchmarks as standardized evaluation instruments enable measuring certain capabilities and provide comparisons across systems~\cite{Raji:2021Everything}. Because benchmarks serve as shared standards, their results are interpreted as indicators of model performance and guide further development toward better performance on benchmark \task{s}~\cite{Orr:2024Sport}.

However, benchmark design is not a trivial endeavor, and without a proper design, benchmarks may fail to reveal the limitations of systems toward progress. Research on the challenges of benchmarking has shown that benchmarks may not necessarily measure the capabilities they claim to measure, \ie they lack construct validity~\cite{Liu:2024Ecbd, Saxon:2024Microscopes}, or they may fail to account for how models are used in real-world contexts, \ie they suffer from ecological validity~\cite{Hardy:2025Marketing, Ethayarajh:2020Utility}. This same issue is applicable to health-related benchmarks as they make assumptions about downstream use cases and how people interact with LLMs in the real-world~\cite{Rajkomar:2026Validity, Raman:2026Assumptions}.
To address such shortcomings, researchers have proposed principled approaches to benchmark design, including frameworks for justifying design choices with validity evidence~\cite{Liu:2024Ecbd}, and evaluations grounded in the use cases and requirements of the people who use models~\cite{Liao:2023Narrowing}.  
Human-centered approaches further explored bringing people's needs and workflows into benchmark design by engaging domain practitioners~\cite{Li:2026Journalism}, community organizations~\cite{Hamna:2026Community}, and end users~\cite{Palani:2026Lexara} in informing tasks and evaluation criteria. 
Taking this human-centered approach is particularly important for personal health informatics benchmarks given the complexity of the process and its inter-related and evolving components or else benchmark evaluations are not aligned with real-world people's needs.
Within this line of research, our work draws on theories and empirical knowledge of \phinfo\ to examine benchmark design.

\section{Methodology}
\label{sec:method}
We examine existing benchmarks to understand whether and how well they establish if LLM systems can support personal health informatics activities. Below, we describe how we identified benchmarks to examine~(\sect{sec:method-selection}) and how we subsequently analyzed different aspects of their designs~(\sect{sec:method-analysis}). 

\subsection{Benchmark Selection}
\label{sec:method-selection}
We used a combination of snowballing \cite{Wohlin:2014Snowballing} and database search to select benchmarks. Both methods are commonly used in identifying relevant literature in Human-Computer Interaction~(HCI) research \cite{Raza:2026Augmented, Cha:2025Collaborative, Wischnewski:2023Trust}. 
We used as seed papers for snowballing: \healthbench{}~\cite{Arora:2025Healthbench}, which provides a comprehensive evaluation of LLMs across health-related scenarios, and \citet{Merrill:2026Wearable}'s work on user-facing interactions with wearable health data, which closely aligns with \phinfo{} activities. These papers also presented differing benchmark design choices that strengthened our snowballing. 
We conducted both backward and forward snowballing to find additional papers. 
We broadened our search by querying databases that index personal health informatics and benchmarking research: ACM DL, PubMed, Scopus and arXiv. We used such search terms as `benchmark', `large language models', `\selftracking' and `personal health' for papers published between December 2022 and \searchEndDate{} (see \appref{app:search-query} for details on our search query).
We used the following selection criteria to narrow from 5,734 papers that came from snowballing and database search. 

\subsubsection{Inclusion and Exclusion Criteria}
\label{sec:method-selection-criteria}
We define personal informatics as the process of collecting information about oneself to address such needs as obtaining self-knowledge~\cite{Li:2010Stage} or instrumenting one's activities and goals~\cite{Epstein:2015Lived}. 
We define a benchmark as a collection of material, namely, \textit{tasks}, \textit{protocols}, and \textit{metrics} where a system performance can be measured on the tasks via metrics following the protocols. Benchmark tasks, protocols, and metrics are designed to demonstrate whether and to what degree systems possess certain capabilities.
Given these definitions, we included papers that evaluated LLMs or LLM-powered systems on (1) tasks framed as an individual making an inquiry involving their self-tracked health data, 
which could be passively sensed (\eg wearable data) or actively reported (\eg\ self-logged records), 
through natural-language interactions, 
and (2) offered material that could be reused to examine capabilities of LLMs or LLM-powered systems. 
We excluded papers that 
(1)~focused on clinically performed tasks such as disease diagnosis or treatment planning given clinical indications and input~(\eg adjusting medication dosage based on symptoms), 
(2)~evaluated general personal health questions without a connection to personal data~(\eg learning about general guidelines for physical activity levels\ycomment{DONE\_FM include a more concrete example if time permits}), 
(3)~only required information such as demographics or general profiles (\eg diet type or family history of health conditions) rather than longitudinal self-tracked data~(\eg food recommendations based on condition-related dietary restrictions\ycomment{DONE\_FM include a more concrete example if time permits}), 
or (4)~primarily emphasized `sensing', \ie the extraction of health markers, typically from raw physiological signals~(\eg deriving heart rate, heart-rate variability, or the presence of arrhythmia from ECG signals). 
We also excluded papers where the evaluation included elements that were too specific to the system being tested, for example, papers that evaluated a fine-tuned model using tasks aligned with its training examples~\cite{Khasentino:2025Sleep}. 

\subsubsection{Screening Procedure}
\label{sec:method-selection-procedure}
We used AI's assistance~(Claude Sonnet 4.6) to screen papers given a prompt that included the inclusion and exclusion criteria above and asked for one of the following labels for each paper~(\appref{app:screening-prompt}): `included', `excluded', or `not sure'. We first validated the AI assistant's performance by comparing its output against a subset of papers manually and independently reviewed for relevance given our selection criteria~(\sect{sec:method-selection-criteria}) before
applying it to the full set of papers. On the validation set of \nSnowball\ papers, 
the AI screener retained all papers that we had identified as relevant, correctly excluded papers that we had determined to be irrelevant, and classified fewer than 1\% of the papers as `not sure'. It did not exclude any papers that we had previously identified for inclusion. It produced this output only based on title and abstract information. We were thus able to perform the screening at the scope of title and abstract over the full set of papers at reasonable cost with no loss in performance. 
%
AI screening returned \nAIlabeledIncluded\ papers with the label `included', which we manually reviewed. 
When a paper text did not include sufficient information to determine its relevance, we examined the associated benchmark material if published. We excluded papers when such information was unavailable. 
Through this procedure we arrived at the final corpus of \nBenchmarks\ benchmarks summarized in~\tab{tab:benchmarks}. We use the short name we have given each benchmark in the table in the remainder of the text. \fatemeh{a decision we may want to make: for the order of papers in table is it better to list based on method narrative (the order we found them?) or number of their tasks}

\begin{table*}[t]
\centering
\caption{Overview of benchmarks included in our corpus. Availability indicates whether benchmark \taskeg{s} are publicly available. Re-usability indicates whether the benchmark material~(\ie \taskeg{s}, protocols, and metrics) can be reused to evaluate new systems.
}
\label{tab:benchmarks}

\small
\renewcommand{\arraystretch}{1.25}
\setlength{\tabcolsep}{7pt}

\begin{tabularx}{\textwidth}{
@{}
p{0.16\textwidth}
>{\centering\arraybackslash}p{0.2\textwidth}
X
>{\centering\arraybackslash}p{0.1\textwidth}
>{\centering\arraybackslash}p{0.1\textwidth}
@{}
}
\toprule
\textbf{Benchmark} &
\textbf{\# \TaskEg{s}} &
\textbf{Domain} &
\textbf{Availability} &
\textbf{Re-usability} \\
\midrule

\cgmspeak~\cite{Cui:2026Cgm}
& 4,180
& CGM analysis; diabetes management
& Yes
& No \\

\addlinespace[3pt]
\dmbench~\cite{Cardei:2025Dmbench}
& 360,600
& Personalized diabetes management
& No
& Yes \\

\addlinespace[3pt]
\healthbench~\cite{Arora:2025Healthbench}
& 5,000
& Broad healthcare
& Yes
& Yes \\

\addlinespace[3pt]
\lifeagentbench~\cite{Tian:2026Lifeagentbench}
& 22,573
& Lifestyle health data analysis
& Yes
& Yes \\

\addlinespace[3pt]
\llmcgm~\cite{Healey:2024LLMcgm}
& 30
& CGM analysis; diabetes management
& Yes
& No \\

\addlinespace[3pt]
\perqa~\cite{Christmann:2026Perqa}
& 17,632
& Heterogeneous personal data analysis
& Yes
& Yes \\

\addlinespace[3pt]
\pha~\cite{Heydari:2025Anatomy}
& 50
& Multi-agent health coaching
& No
& No \\

\addlinespace[3pt]
\phia~\cite{Merrill:2026Wearable}
& 4,172
& Wearable health data analysis
& Yes
& No \\

\bottomrule
\end{tabularx}
\Description{
A table summarizing eight benchmarks included in the review corpus. 
Columns report the benchmark name, number of task examples, domain, whether the task examples are publicly available, and whether the benchmark materials can be reused to evaluate new systems. 
CGM Speak contains 4,180 task examples in CGM analysis and diabetes management; its tasks are publicly available, but its benchmark materials are not reusable. 
DM-Bench contains 360,600 task examples in personalized diabetes management; its tasks are not publicly available, but its benchmark materials are reusable. 
HealthBench contains 5,000 task examples covering broad healthcare; its tasks are publicly available and its benchmark materials are reusable. 
LifeAgentBench contains 22,573 task examples for lifestyle health data analysis; its tasks are publicly available and its benchmark materials are reusable. 
LLM-CGM contains 30 task examples in CGM analysis and diabetes management; its tasks are publicly available, but its benchmark materials are not reusable. 
PerQA contains 17,632 task examples involving heterogeneous personal data analysis; its tasks are publicly available and its benchmark materials are reusable. 
PHA contains 50 task examples for multi-agent health coaching; its tasks are not publicly available and its benchmark materials are not reusable. 
PHIA contains 4,172 task examples for wearable health data analysis; its tasks are publicly available, but its benchmark materials are not reusable.
}
\end{table*}

\subsection{Benchmark Analysis}
\label{sec:method-analysis}
We extracted information on `\tasks' that benchmarks evaluated along with their associated `\evalsetting' in summary notes and tables that we used in further analysis. \textit{\Tasks} are classes of test cases, with each individual test case referred to as a `\onetaskex'. \textit{\Evalsetting}s include information on the creation and verification of tasks along with protocols and metrics of system evaluation and the respective verification of these elements.
In addition to the full text of each paper of the corpus, we reviewed 
supplementary material, code repositories, and the \textit{dataset} of \taskex\ (when available).
For benchmarks that covered a broad range of \tasks, we only analyzed the subset that aligned with our criteria~(see \sect{sec:method-selection-criteria}). 
This process was primarily done manually, except for \healthbench~\cite{Arora:2025Healthbench}, 
where an initial manual examination indicated that \taskex{} meeting our inclusion criteria were sparse within the relatively large dataset ($n=5000$). We designed a classification prompt~(\appref{app:healthbench-prompt}) with Claude Sonnet 4.6 to label the \taskex{} and verified the classification performance on 500 \taskex{}. 

We conducted reflexive thematic analysis \cite{Braun:2006Thematic, Braun:2019Reflexive} to characterize the design space of benchmarks in terms of what they were testing~(\ie \tasks) and how~(\ie \evalsetting s), to then examine how well these design choices were aligned with needs in \phinfo\ activities, and to subsequently identify opportunities for improvements. 
In this process, we drew on our expertise in \phinfo{} research and our research on people's lived experience with \selftracking\ tools. 

We combined deductive and inductive coding to analyze benchmark \tasks\ and their underlying structures. Deductive coding was informed by information needs identified in \phinfo\ research \cite{Li:2011Understanding, Rey:2023Investigating, Sefidgar:2025Thesis}, such as questions individuals ask about how a health or behavior metric varies over time, or how different factors and a health metric are related. Inductive coding occurred as we reviewed benchmark \taskex\ and new details emerged that were not previously captured. 
We iteratively grouped codes into themes representing categories of capabilities measured in benchmarks. Each category captured a specific type of capability assessed by benchmarks and, where applicable, its underlying structural components and subtypes. 
We leveraged AI assistance in reviewing \task{s} as these were numbered in the thousands. We developed codes based on the review of a small initial subset of \task{s}, then created classification prompts for Claude Sonnet 4.6 to apply the codes over the larger set of \task{s}. We iteratively refined the prompts to reach desirable accuracy in a manually coded subset of \task{s} that was expanded in each refinement iteration. We also refined codes as needed if the iteration surfaced new insights. We used this AI-assisted coding when identifying categories of tasks and when coding the structural components of the task categories with over 5000 task examples. For the former, we reached 90\% accuracy in 25\% of \taskeg{s} for six \task\ categories. For the latter, we were at 85\% accuracy in 10\% of \taskeg{s}. \appref{app:task-mapping-prompt} and \appref{app:variation-mapping-prompt} provide the details of the coding prompts. 

We similarly examined design decisions made in benchmark \evalsetting s to demonstrate LLM capabilities. 
For example, we examined the type and temporal scope of personal data provided as task data and how those data were collected or synthetically generated. We iteratively grouped related codes into conceptual design dimensions and identified the values observed within each dimension. 
Examples of these dimensions include `method of task creation', `task data properties', and `response generation settings'. 

Some of the papers of our corpus were included only in our analysis of \taskex\ and data characteristics but excluded with respect to dimensions related to `response generation and scoring' because their evaluation setup was too system-specific or otherwise not reusable for scalable evaluation. For example, \pha{}~\cite{Heydari:2025Anatomy} evaluated their multi-agent system using human raters, without providing a scalable scoring method that could be reused to evaluate new systems. 
In addition to \pha{}, LLM-CGM~\cite{Healey:2024LLMcgm}, CGM Speak~\cite{Cui:2026Cgm}, and \phia{}~\cite{Merrill:2026Wearable} were excluded from the analysis of `response generation and scoring'. 

We analyzed the resulting characterizations of \phinfo{} benchmark \tasks\ and \evalsetting s to identify potential gaps in existing benchmarking practices. Throughout this process, the research team documented observations that emerged during discussions, such as the dominance of analytical tasks and the limited representation of tasks concerning earlier stages of \selftracking. 
We subsequently conducted a thematic analysis of the combined notes using affinity diagramming \cite{Braun:2006Thematic} to organize related observations into higher-level themes. Examples of the final themes include design-related gaps and reporting issues in benchmark papers.
\section{Results}
\label{sec:results}
We organize benchmark design space along two major subspaces of \textit{\task}s and \textit{\evalsetting}s consistent with the definition of benchmarking~(\sect{sec:method-selection-criteria}). We first offer the characterization of these subspaces~(\sect{sec:result-design-space}). We then use this characterization as a language to map out the current landscape of benchmark design for \phinfo\ activities~(\sect{sec:results-landscape}). We next demonstrate what is missing but needed in existing benchmarks~(\sect{sec:results-gaps}). 
Briefly, we find 
benchmark \task{s} and \evalsetting{s} do not properly incorporate domain knowledge, 
there is lack of representation for activities in stages of tracking other than data analysis, 
and unrealistic assumptions and lack of consideration for end-to-end tracking experience limit the ecological validity of benchmarking.
\subsection{Characterization of Personal Health Informatics Benchmarks}
\label{sec:result-design-space}
Benchmarking is the practice of examining system capabilities over a set of test cases~(\ie the \textit{\taskeg}s under \textit{\task} categories), within a testing ground~(\ie an \textit{\evalsetting}). The latter is defined by specific protocols for creating and verifying \taskeg{s} and their input, testing procedures, and performance metrics. Below, we lay out the types and dimensions that can identify \task{s} and \evalsetting{s} in existing \phinfo\ benchmarks.
\subsubsection{Characterization of Benchmark Tasks}
\label{sec:result-design-space-task}
\begin{figure*}[t]
    \centering
    \includegraphics[width=\textwidth]{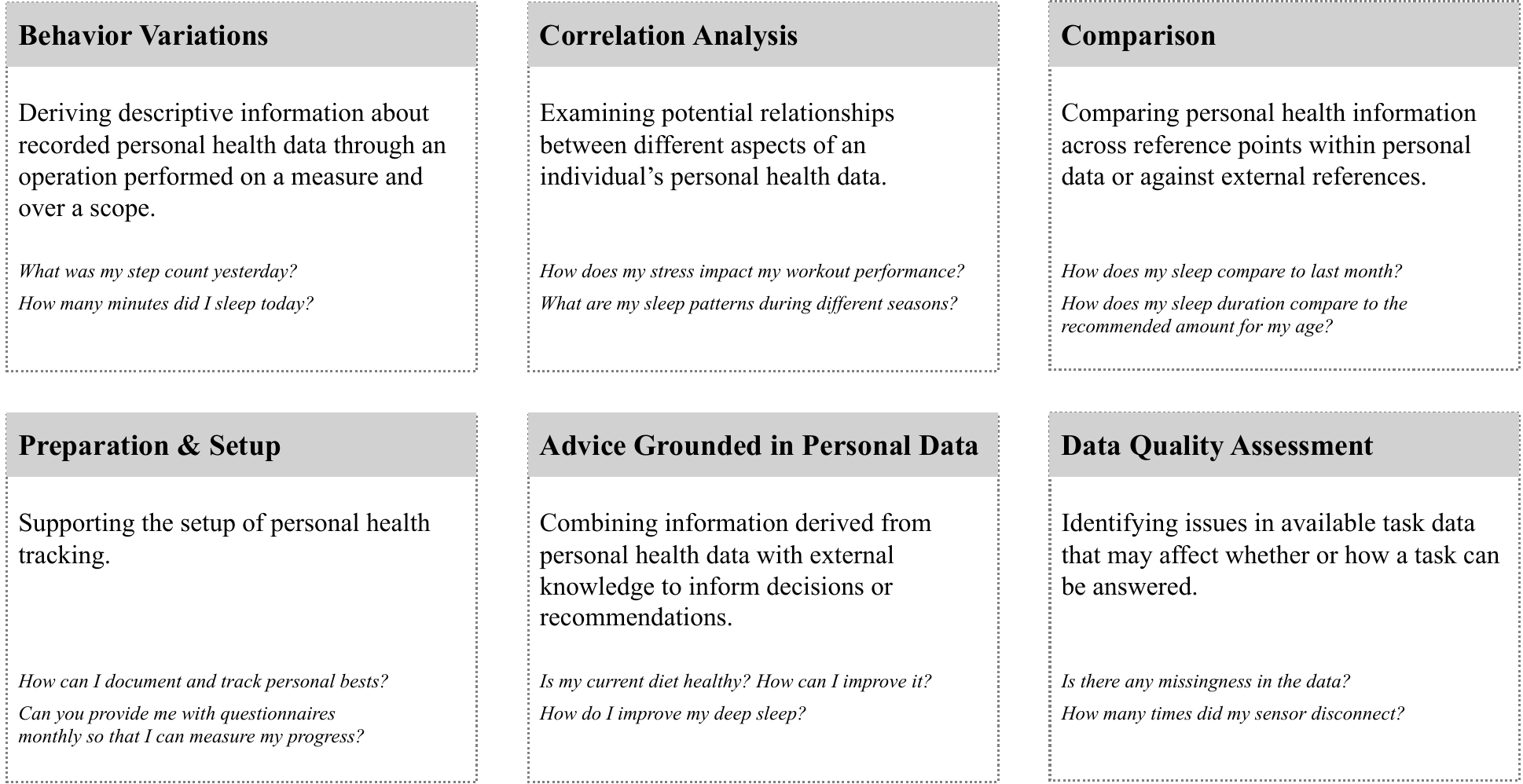}
    \caption{
            Six categories of benchmark \task{s} identified across the \nBenchmarks\ benchmarks in our corpus. Each card names a category, defines the capability of the \sut, and gives two \taskeg{s} from the benchmarks.}
    \Description{A six-panel diagram summarizing categories of benchmark tasks. 
    Behavior Variations covers deriving descriptive information about recorded personal health data through an operation performed on a measure over a scope; example questions ask ``What was my sleep count yesterday?'' and ``How many minutes did I sleep today?'' 
    Correlation Analysis covers examining potential relationships between aspects of an individual's personal health data; examples ask how stress impacts workout performance and about sleep patterns during different seasons. 
    Comparison covers comparing personal health information across reference points within personal data or against external references; examples compare sleep with the previous month and sleep duration with recommended amounts for the user's age. 
    Preparation and Setup covers supporting the setup or revision of personal health tracking and data collection practices; examples ask how to document and track personal bests and request monthly questionnaires to measure progress. 
    Advice Grounded in Personal Data covers combining information derived from personal health data with external knowledge to inform decisions or recommendations; examples ask how to improve sleep based on poor sleep scores and how to improve deep sleep. 
    Data Quality Assessment covers identifying issues in available task data that may affect whether or how a task can be answered; examples ask whether there are missing values in the data and how many times a sensor disconnected.}
    \label{fig:tasks_char}
\end{figure*}

We identified six categories of benchmark \tasks\ presented below and summarized in \fig{fig:tasks_char}.
Each category represents a capability in the \sut\ and is often itself characterized by distinct structures.

\paragraph{\textbf{\taskBehaviorVar.}} 
\Task{s} that 
can be represented as ([\textit{operation}], [\textit{measure}], [\textit{scope}]), where 
`operation' is one of `retrieving a specific value or set of values', hereafter referred to as `value lookup'~(\eg \qt{How long was my run yesterday?}~\cite{Merrill:2026Wearable}), `performing an aggregation over a set of records'~(\eg average, sum), or `identifying a trend across the records'~(\eg \qt{Is there a trend in my blood glucose data over last week?}~\cite{Cui:2026Cgm}) \ycomment{TO-DO\_FM change the example}; 
%
`measure' is the recorded value of an attribute of a behavior or state, such as the duration or speed of a workout session, the occurrence or severity of symptoms of a chronic condition, or blood glucose levels of glycemic activity; 
`scope' specifies the subset of records over which the operation is performed. It can be global~(\ie across all records with no filtering), or can identify a form of filtering, such as a time window (\eg `past month') or a numeric threshold (\eg `workout sessions longer than 30 minutes'). 
For example, \qt{what's the total distance I ran last month?~\cite{Christmann:2026Perqa}} is a \taskbehaviorvar\ task, where `sum' is the operation applied to `distance of run' as the measure over the scope of `last month'.
Similarly, \qt{How many minutes did I sleep today?~\cite{Merrill:2026Wearable}} is a \taskbehaviorvar\ task with `value lookup' as the operation that retrieves `duration of sleep' as the measure within `today' as the scope. 
%
We have listed the variety of operations, measures, and scope values in \appref{app:variation-details}.

\paragraph{\textbf{\taskCorrelation.}} 
\Tasks\ that 
examine potential relationships between different aspects of an individual's personal health data~(\eg \qt{How does my sleep duration relate to my daily steps?}~\cite{Merrill:2026Wearable}). Correlation \task{s} can vary in the temporal alignment of their factors that are either concurrent (\eg same-day relation of stress levels and workout performance) or delayed (\eg whether poor sleep determines blood glucose within three days). Sometimes, behavior metrics are correlated with time~(\eg \qt{What are my sleep patterns during different seasons?}~\cite{Merrill:2026Wearable}). The time variable is either a named period, such as weekdays~(\eg \qt{How did my resting heart rate differ on weekends \vs\ weekdays last month?}), or an event~(\eg \qt{Can you tell me how my sleep quality was affected by daylight savings?}). 

\paragraph{\textbf{\taskComparison.}} 
\Task{s} that 
consider differences between personal health information at two or more reference points~(\eg \qt{How does my sleep compare to last year?}~\cite{Merrill:2026Wearable}). \taskComparison{s} may be made within an individual's own data~(\eg comparing a behavior metric at two points in time or two different contexts) or against external reference values, such as population norms and demographics standards or common guidelines~(\eg \qt{How does my sleep duration compare to the recommended amount for my age?}~\cite{Merrill:2026Wearable}). 

\paragraph{\textbf{\taskDataQual.}}
\Task{s} that examine the identification of potential issues in task data, such as missing values~(\eg \qt{How many times did my sensor disconnect?}~\cite{Healey:2024LLMcgm}). 

\paragraph{\textbf{\taskPItoAdvice.}}
\Task{s} that assess combining external knowledge with personal health information to inform data-grounded decision making (\eg \qt{Is my current diet healthy? How can I improve it?}~\cite{Merrill:2026Wearable}). 
These \task{s} typically involve contextualizing information from the individual's data using external knowledge~(\eg clinical guidelines) to provide a recommendation or suggested action.

\paragraph{\textbf{\taskPrep.}}
\Task{s} that 
involve an individual asking a \sut\ for data collection tools~(\eg a tracking table for recording blood pressure~\cite{Arora:2025Healthbench} or questionnaires for \qt{measuring progress}~\cite{Merrill:2026Wearable}) or for advice on deciding the details of their \selftracking~(\eg what to track and how often)~\cite{Arora:2025Healthbench}.

\subsubsection{Characterization of \EvalSetting{s}}
\label{sec:result-design-settings}

\begin{figure*}[t]
    \centering
    \includegraphics[width=\textwidth]{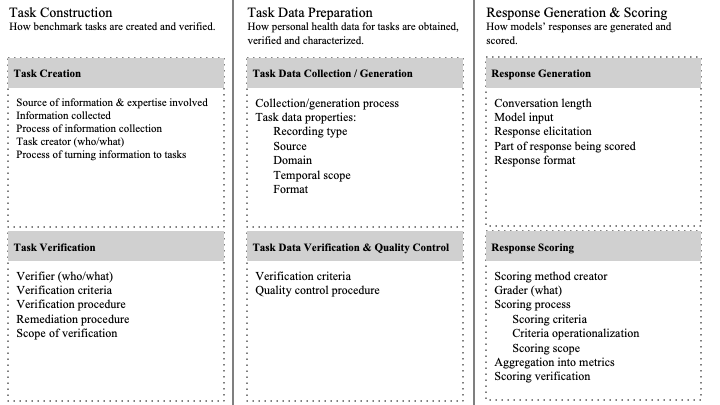}
    \caption{Six dimensions that characterize benchmark evaluation settings, along with the sub-dimensions of design choices within each. Task Creation describes how benchmark tasks are informed and prepared, and Task Verification describes how those tasks are checked before inclusion. Task Data Collection/Generation describes how the personal health data paired with tasks are obtained and what properties they have, and Task Data Verification and Quality Control describes how task data are checked and prepared for use in the benchmark. Response Generation characterizes how the \sut\ responds to tasks, and Response Scoring describes how the \sut’s output is graded and aggregated into metrics, and how the scoring itself is verified. Response Generation and Response Scoring apply only to benchmarks whose material can be reused to evaluate other systems. The values observed for each sub-dimension are reported in \appref{app:settings-details} (\tab{tab:task-creation-verification}--\tab{tab:response-generation-scoring}).}
    
    \Description{Diagram characterizing benchmark evaluation settings across three areas: Task Construction, Task Data Preparation, and Response Generation and Scoring. Under Task Construction, the figure includes Task Creation and Task Verification. Task Creation is characterized by the source of information and expertise involved, the information collected, the process of information collection, the task creator (who or what created the task), and the process of turning collected information into tasks. Task Verification is characterized by the verifier (who or what verifies the task), verification criteria, verification procedure, remediation procedure, and scale of verification. Under Task Data Preparation, the figure includes Task Data Collection/Generation and Task Data Verification and Quality Control. Task Data Collection/Generation is characterized by the collection or generation process and by task data properties, including recording type, source, domain, temporal scope, and format. Task Data Verification and Quality Control is characterized by verification criteria and the quality control procedure. Under Response Generation and Scoring, the figure includes Response Generation and Response Scoring. Response Generation is characterized by conversation length, model input, response elicitation, the part of the response that is scored, and response format. Response Scoring is characterized by the scoring method creator, the grader (who or what performs grading), the scoring process, including scoring criteria, criteria operationalization, and scoring scope, as well as aggregation into metrics and scoring verification.}
    \label{fig:char-settings}
\end{figure*}

We identify benchmark \evalsetting{s} along six dimensions 
of creating and verifying \task{s} and their input data as well as procedures and metrics for evaluating \sut's performance~(\fig{fig:char-settings}). 
Each dimension is further characterized using sub-dimensions that describe the design choices involved in that dimension~(see full details in \appref{app:settings-details}).

\paragraph{\textbf{Task Creation.}} This \settingsDIM\ characterizes the process or authoring strategy used to produce the benchmark \taskex{}. We describe this process in terms of 
the source of information and expertise involved in task creation~(\eg benchmark developers, domain experts), 
the information collected from those sources~(\eg sample user queries, query templates), 
the method used to collect that information~(\eg surveys, red-teaming), 
the entity who created \taskeg{s}~(\eg benchmark designers, LLMs), 
and the process by which \taskeg{s} were created based on the collected information~(\eg if information was directly used without modifications, or if there were any intermediate transformations). 

\paragraph{\textbf{Task Verification.}} This \settingsDIM\ captures information on how \taskex{} are checked for quality before inclusion in a benchmark. We describe this process in terms of 
the entity who performed the verification~(\eg human experts or LLMs), 
aspects of the \taskeg{s} that were verified~(\eg linguistic quality or representativeness of intended capabilities), 
verification method~(\eg scoring \taskeg{s} against predefined criteria), 
remediation process when verification issues were identified~(\eg refinement of \taskeg{s}), 
and the scope of verification~(\eg review of all or a subset of \taskeg{s}). 

\paragraph{\textbf{Task Data Collection/Generation.}} This \settingsDIM\ concerns the preparation of the personal health data provided as input alongside \taskeg{s} in benchmarks, hereafter referred to as \textit{task data}. 
We describe this process in terms of 
the method for obtaining task data~(\eg synthetically generated by neural networks or collected from real individuals in a user study), 
across types~(\eg passively sensed or actively reported), 
domains~(\eg physical activity, medication usage), 
and temporal scopes~(\eg one month), 
as well as the format for presenting the data (\eg structured tabular data or unstructured text).

\paragraph{\textbf{Task Data Verification and Quality Control.}} This \settingsDIM{} characterizes ways the utility of task data is established or verified for \task{s} a benchmark targets. We describe this process in terms of 
the criteria~(\eg realism) 
and the process used for verification and quality control~(\eg injecting missing values to mirror sporadic patterns often observed in real data).

\paragraph{\textbf{Response Generation.}} This \settingsDIM\ identifies the process by which benchmark \taskeg{s} are presented to the \sut\ to elicit the responses that are subsequently scored in assessing the system's performance. We describe this process in terms of 
the number of conversational turns involved~(\eg single- or multi-turn),
the input provided to the \sut~(\eg a \taskeg\ together with its task data),
response elicitation procedure~(\eg prompting the system with the \taskeg\ and its task data), 
the part of the generated conversation considered for scoring~(\eg the final generated response or the full conversations) 
and the expected response format (\eg free-form text \vs\ a numeric or one-word response).

\paragraph{\textbf{Response Scoring.}} This \settingsDIM\ is about how responses generated by the \sut\ are scored, and how scores are aggregated into performance reports, and how the scoring is verified. We describe this process in terms of
who defines the scoring method~(\eg domain experts or end user insights),
the entity who performs the scoring~(\eg an LLM grader or a rule-based grader),
the scoring rubric defined as the criteria for assessing responses~(\eg accuracy or clarity), the operationalization of the criteria~(\eg free-form text description, potentially with positive and negative examples) as well as the procedure and scope of applying the criteria~(\eg if criteria is applied benchmark-wide or to specific \task{s}), 
how response scores per \taskeg\ are aggregated into reporting metrics~(\eg\ mean of scores), 
and how the scoring method is verified. 
Verification  is designed in terms of 
the aspect being verified, 
the entity who performs the verification, 
and the process~(\eg independent expert ratings can verify the reliability of LLM-grader in applying the rubric).

\subsection{Landscape of Personal Health Informatics Benchmarks}
\label{sec:results-landscape}
We describe the design choices of benchmarks in our corpus using the language established in \sect{sec:result-design-space} around the categories of benchmark \tasks{} and the \settingsDIM{s} of \evalsetting{s}. The characterization of each benchmark is provided in the supplementary material, with \cgmspeak{} included in \appref{app:char-cgmspeak} as an example.
\subsubsection{Landscape of Benchmark Tasks}
\label{sec:results-landscape-tasks}

\begin{figure*}[t]
    \centering
    \includegraphics[width=\textwidth]{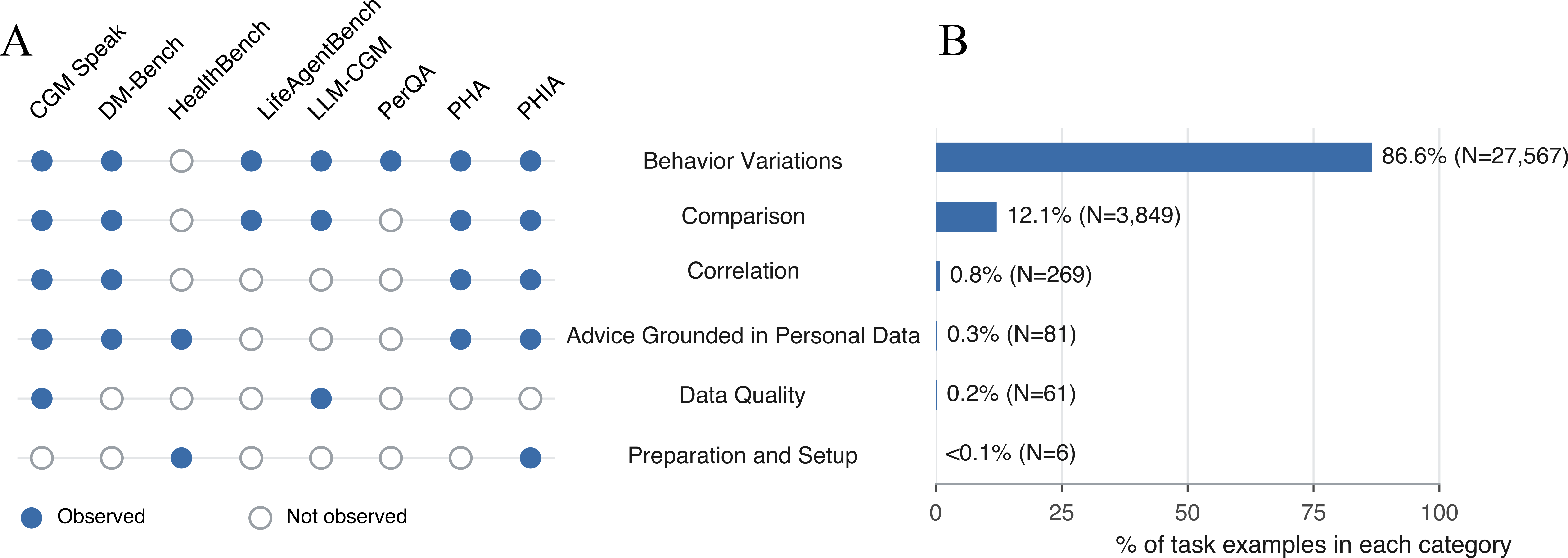}
    \caption{ Landscape of benchmark \tasks{}.
    (A)~Presence of each task category across the \nBenchmarks{} benchmarks in the corpus. 
    Filled circles indicate that a category was observed, and open circles indicate that it was not.
    (B) Distribution of analyzed \taskeg{s} across the categories of benchmark \tasks{} described in \sect{sec:result-design-space-task} (n = \nPHIBenchmarkTasks). (A) and (B) share the task category labels on the vertical axis in the middle.
    \taskBehaviorvar\ dominated the distribution (\nBehaviorVar), whereas \tasks{} concerning preparation for tracking and the quality of personal health data were sparse (\nPrep\ and \nDataQual, respectively). For \lifeagentbench{}, \taskeg{s} with multiple queries were split and each query was counted in its own category (\eg a \taskeg{} asking both how much higher the step count was on a date than on the previous day and which food was most frequent counts as one comparison and one variation).
    \dmbench{} and \pha{} are not represented in this distribution because the full datasets of their \taskeg{s} are not publicly available; as described in \sect{sec:method-analysis}, we categorized their \tasks{} using the descriptions and examples in their papers.}
    
    \Description{
    Two-panel figure showing the distribution and benchmark coverage of six categories of benchmark tasks. Panel A is a horizontal bar chart with a logarithmic x-axis showing the percentage of analyzed task examples in each category. Behavior Variations is the dominant category at 94.3\%, followed by Comparison at 4.2\%, Correlation at 1.0\%, Advice Grounded in Personal Data at 0.3\%, Data Quality at 0.2\%, and Preparation and Setup at less than 0.1\%. Panel B is a matrix showing whether each of these six task categories was observed across eight benchmarks: CGM Speak, DM-Bench, HealthBench, LifeAgentBench, LLM-CGM, PerQA, PHIA, and PHA. Rows correspond to the six task categories in the same order as Panel A. Filled blue circles indicate an observed category and open circles indicate a category that was not observed. Behavior Variations is observed in all benchmarks except HealthBench. Comparison is observed in CGM Speak, LifeAgentBench, LLM-CGM, PHIA, and PHA. Correlation is observed in CGM Speak, DM-Bench, PHIA, and PHA. Advice Grounded in Personal Data is observed in CGM Speak, DM-Bench, HealthBench, PHIA, and PHA. Data Quality is observed in CGM Speak and LLM-CGM. Preparation and Setup is observed in HealthBench and PHA.
    }
    \label{fig:task-landscape}
\end{figure*}

\fig{fig:task-landscape} shows the distribution of \taskeg{s} across the \task{} categories described in \sect{sec:result-design-space-task}, along with a coverage map of where (\ie which benchmark) these \task{s} were observed. \taskBehaviorvar\ category is by far the most common, accounting for \nBehaviorVar\ of analyzed \taskex{} and appearing in \nCorpusBehaviorVar\ of the \nBenchmarks\ benchmarks in our corpus. More broadly, data analysis \task{s}~(\ie `\taskBehaviorVar', `\taskComparison', and `Correlation') represent 99.5\% of the \taskeg{s}. The remaining categories~(\ie `\taskPrep', `\taskPItoAdvice', and `\taskDataQual') are rarely observed. 

\begin{figure*}[t]
    \centering
    \includegraphics[width=\textwidth]{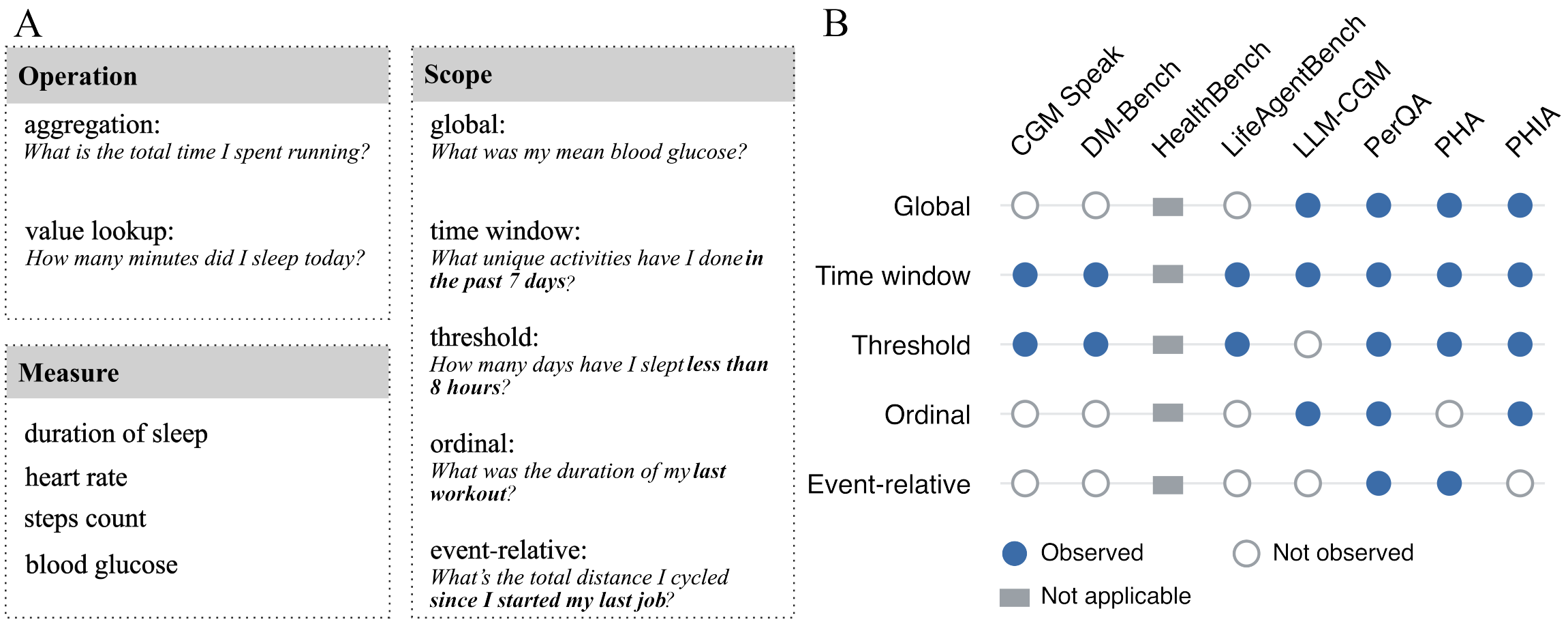}
    \caption{
    Structure and observed values of \qt{\taskBehaviorVar} \taskex{}.
    (A) \taskex{} are characterized using three components: operation, measure, and scope. Observed operations include aggregation and value lookup; measures represent recorded behavioral or physiological attributes; and scopes specify which records are considered, including global, time-window, threshold, ordinal, and event-relative scopes. Representative \taskex{} illustrate each operation and scope type.
    (B) Presence of the five scope types across benchmarks in the corpus. Filled circles indicate that a scope type was observed, open circles indicate that it was not observed, and gray squares indicate benchmarks for which no \qt{\taskBehaviorVar} task examples were available to characterize, so the scope types are marked as not applicable.
    }
    \Description{
    Two-panel figure characterizing the structure of Behavior Variations task examples and the scope types represented across benchmarks. Panel A shows that a Behavior Variations task example consists of an operation, a measure, and a scope. The observed operation types are aggregation and value lookup. Aggregation applies a calculation over records, illustrated by the question ``What is the total time I spent running?'' Value lookup retrieves a recorded value, illustrated by ``How many minutes did I sleep today?'' Example measures include duration of sleep, heart rate, step count, and blood glucose. Five scope types are shown. Global applies no additional restriction, illustrated by ``What was my mean blood glucose?'' Time window restricts records to a specified period, illustrated by ``What unique activities have I done in the past 7 days?'' Threshold restricts records according to a condition, with examples including ``How many days have I slept less than 8 hours?'' and ``How many days did I consume food in the Vegetables and Fruits category?'' Ordinal selects records according to their ordering or rank, illustrated by ``What was the duration of my last workout?'' Event-relative restricts records relative to an event, illustrated by ``What's the total distance I cycled since I started my last job?'' Panel B is a matrix showing whether the five scope types were observed in eight benchmarks: CGM Speak, DM-Bench, HealthBench, LifeAgentBench, LLM-CGM, PerQA, PHIA, and PHA. Global scope is observed in CGM Speak, LifeAgentBench, LLM-CGM, PerQA, PHIA, and PHA, but not in DM-Bench; it is not applicable to HealthBench. Time-window scope is observed in all applicable benchmarks and is not applicable to HealthBench. Threshold scope is observed in CGM Speak, DM-Bench, LifeAgentBench, PerQA, and PHA; it is not observed in LLM-CGM or PHIA and is not applicable to HealthBench. Ordinal scope is observed in LLM-CGM, PerQA, and PHA; it is not observed in CGM Speak, DM-Bench, LifeAgentBench, or PHIA and is not applicable to HealthBench. Event-relative scope is observed in PerQA and PHIA; it is not observed in CGM Speak, DM-Bench, LifeAgentBench, LLM-CGM, or PHA and is not applicable to HealthBench. Filled circles denote observed values, open circles denote values not observed, and gray squares denote cases where the characterization is not applicable.
    }
    \label{fig:variation}
\end{figure*}

\begin{figure*}[t]
    \centering
    \includegraphics[width=\textwidth]{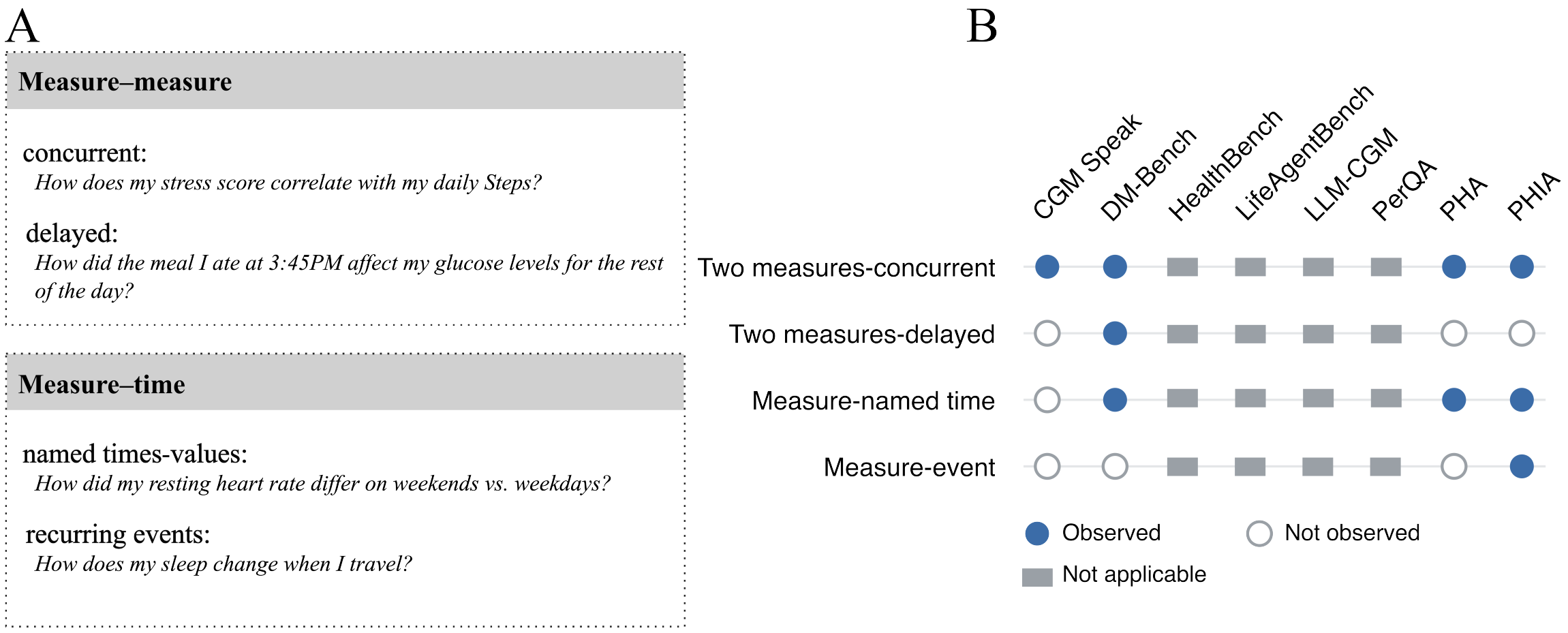}
    \caption{
    Types and benchmark coverage of \qt{\taskCorrelation} \tasks{}.
    (A) Two forms of correlation analysis are observed: measure--measure relationships, which can be concurrent or delayed, and measure--time relationships, where measures are related to named times or recurring events. Representative \taskex{} illustrate each subtype.
    (B) Presence of the four correlation subtypes across benchmarks in the corpus. Measure–measure \task{s} mostly did not specify a temporal alignment between the two measures, implying a concurrent relationship (four benchmarks), whereas delayed relationships were sparse (only \dmbench). Filled circles indicate that a subtype was observed, open circles indicate that it was not observed, and gray squares indicate benchmarks for which no Correlation Analysis task examples were available to characterize, so the subtype is marked as not applicable.
    }
    \Description{
    Two-panel figure characterizing Correlation Analysis tasks and their coverage across eight benchmarks. Panel A shows two forms of correlation analysis. Measure--measure analysis examines relationships between two personal health measures. Concurrent relationships consider measures aligned in time, illustrated by the question ``How does my Stress Score correlate with my daily Steps?'' Delayed relationships consider whether one measure is related to a later value of another measure, illustrated by ``How did the meal I ate at 3:45PM affect my glucose levels for the rest of the day?'' Measure--time analysis examines relationships between a personal health measure and time. Named-times-values relationships compare values across named periods, illustrated by ``How did my resting heart rate differ on weekends vs. weekdays last month?'' Measure--event relationships examine a measure in relation to recurring events, illustrated by ``How does my sleep change when I travel?'' Panel B is a matrix showing whether four correlation subtypes were observed in CGM Speak, DM-Bench, HealthBench, LifeAgentBench, LLM-CGM, PerQA, PHIA, and PHA. Two measures--concurrent is observed in CGM Speak, DM-Bench, PHIA, and PHA; it is marked not applicable for HealthBench, LifeAgentBench, LLM-CGM, and PerQA. Two measures--delayed is observed only in DM-Bench; it is not observed in CGM Speak, PHIA, or PHA and is not applicable to HealthBench, LifeAgentBench, LLM-CGM, or PerQA. Measure--named time is observed in PHIA and PHA; it is not observed in CGM Speak or DM-Bench and is not applicable to HealthBench, LifeAgentBench, LLM-CGM, or PerQA. Measure--event is observed only in PHA; it is not observed in CGM Speak, DM-Bench, or PHIA and is not applicable to HealthBench, LifeAgentBench, LLM-CGM, or PerQA. Filled blue circles indicate observed subtypes, open circles indicate subtypes not observed, and gray squares indicate cases where the subtype was not applicable.
    }
    \label{fig:correlation}
\end{figure*}

\begin{figure*}[t]
    \centering
    \includegraphics[width=\textwidth]{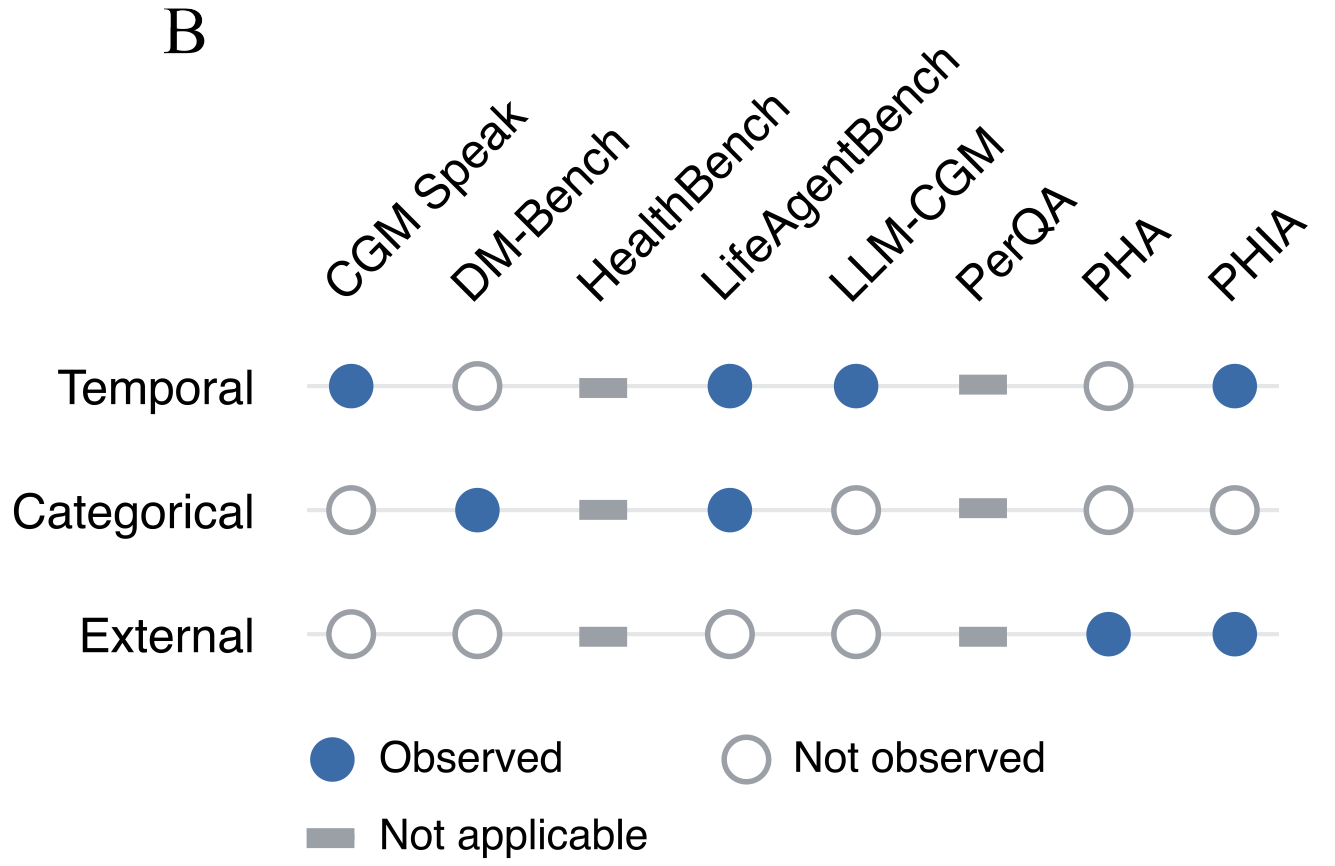}
    \caption{
    Reference points and benchmark coverage of \qt{\taskComparison} \tasks{}.
    (A) Comparison \tasks{} consider differences between personal health information at two or more reference points. Comparisons may be made within an individual's own data, using temporal or categorical reference points, or against external reference values, such as population norms, demographic standards, or clinical guidelines. Representative \taskex{} illustrate each type.
    (B) Presence of the three reference-point types across benchmarks in the corpus. Filled circles indicate that a reference-point type was observed, open circles indicate that it was not observed, and gray squares indicate benchmarks with no \taskComparison{} \taskex{} to characterize.
    }
    \Description{
    Two-panel figure characterizing reference points in Comparison tasks and their coverage across eight benchmarks. Panel A shows three types of reference points. Temporal comparisons consider differences between personal health information at different points in time, illustrated by ``How does my sleep compare to a year ago?'' Categorical comparisons consider differences between personal health information in two different contexts, illustrated by ``Which activity burned more calories: Yoga or Walk?'' External-reference comparisons compare personal health information against external reference values, such as population norms, demographic standards, or common guidelines, illustrated by ``How does my sleep compare to the recommended sleep duration for someone in my age group?'' Panel B shows whether these three reference-point types were observed in CGM Speak, DM-Bench, HealthBench, LifeAgentBench, LLM-CGM, PerQA, PHIA, and PHA. Temporal reference points are observed in CGM Speak, DM-Bench, LifeAgentBench, LLM-CGM, and PHA, and are not observed in PHIA. Categorical reference points are observed only in LifeAgentBench. External reference values are observed in PHIA and PHA. HealthBench and PerQA have no Comparison task examples to characterize. Filled blue circles indicate observed reference-point types, open circles indicate types not observed within benchmarks containing Comparison task examples, and gray squares indicate benchmarks with no Comparison task examples to characterize.
    }
    \label{fig:comparison}
\end{figure*}

Information on values we observed for `\taskBehaviorVar', `\taskCorrelation', and `\taskComparison' \task{s} respectively appear in Figures~\ref{fig:variation}, \ref{fig:correlation}, and \ref{fig:comparison}.
Here, we highlight major observations with respect to these \task{s}. We were able to observe limited variations for the other categories of \task{s}~(\ie `\taskDataQual', `\taskPItoAdvice', `\taskPrep') because of their minimal coverage by the benchmarks. We point out notable patterns we did observe around these \task{s} below. 

\taskBehaviorvar\ \taskeg{s} with aggregation operations were more common in benchmarks which included this \task\ category. 
Measures in \taskbehaviorvar\ ranged across various domains of general \wb, including sleep, physical activity, cardiovascular and glycemic activity, meals, and stress. 
Scopes that filtered personal data records in \taskbehaviorvar\ \taskeg{s} did so based on time windows~(\eg \qt{What was the longest workout session \textbf{last month}?}~\cite{Christmann:2026Perqa}), thresholds~(\eg \qt{What is my average speed on the treadmill when I run faster than \textbf{44 units}?}~\cite{Merrill:2026Wearable}), ordinal position~(\eg \qt{What was the average speed  \textbf{during my last 5 runs}?}~\cite{Christmann:2026Perqa}), events~(\eg \qt{what was the average speed of my runs \textbf{after my last trip}?}~\cite{Christmann:2026Perqa}), or  combinations of these types. There were two major types of thresholding scopes: meeting a fixed numeric or categorical value~(\eg \qt{What are my treadmill steps on days when I have \textbf{less than 9 hours of REM sleep}?}~\cite{Merrill:2026Wearable}), 
or scoping data using a value that was dynamically derived from the data itself to identify outliers (\ie anomaly detection as in \qt{Tell me about anomalies in my active zone minutes last month.}~\cite{Merrill:2026Wearable}; anomalies are divergence from dynamically defined norm values). 

\Taskeg{s} under \taskcorrelation\ that examined the relationship between two behavior measures often did not explicitly specify temporal alignment of the two measures and assumed concurrent same-day alignment~(\ie made assumptions about both the type and the unit of temporal alignment).
In very few cases, the temporal alignment was explicitly specified~(\eg in case of the \textit{delayed} effect of a meal consumed in the afternoon on glucose levels within the few remaining \textit{hours} of the day). 

Reference points of \taskcomparison\ \taskeg{s} were 
temporal~(\eg \qt{Am I a better cyclist than I was 2 years ago based in my app data?}~\cite{Merrill:2026Wearable}), categorical~(\eg \qt{Which
activity burned more calories: Yoga or Walk?}), and external~(\eg \qt{How does my exercise compare to pregnant women my age?}~\cite{Heydari:2025Anatomy}). Unlike \taskcorrelation, these \taskeg{s} compare specific reference points rather than relationships across multiple records.





The few \taskdataqual\ \taskeg{s} in benchmarks primarily asked about incorrect or missing values in wearable data due to potential system errors~(\eg \qt{Given my CGM data over this week, can you identify moments that were likely incorrect blood glucose values?}~\cite{Cui:2026Cgm}). These were typically isolated queries with no follow-up to examine proper handling of data quality issues in subsequent data analysis. 

\Taskeg{s} under \taskpitoadvice\ mostly captured advice seeking either to make adjustments to exercise, diet, or sleep~(\eg \qt{Should I incorporate more cardio with my current physique?~\cite{Merrill:2026Wearable}}) or to initiate medical interventions~(\eg \qt{My glucose has been very high this month, do I need to see a doctor?}~\cite{Cardei:2025Dmbench}).


Rare cases of \taskprep\ \taskeg{s} were about requesting general advice when starting \selftracking{}, often around tracking tools and artifacts~(\eg{} \qt{Can you provide me with questionnaires monthly so that I can measure my progress?}~\cite{Merrill:2026Wearable}). 

\subsubsection{Landscape of \EvalSetting s}
\label{sec:results-landscape-settings}
We observed diverse choices along dimensions of benchmark design (details in \appref{app:settings-details}). Below, we summarize and highlight the most notable patterns.

Different sources of information informed benchmark task creation. This ranged from existing datasets and interaction logs to input from clinicians and end-users, although the latter was less common with end-users directly consulted in only two benchmarks~\cite{Cui:2026Cgm, Heydari:2025Anatomy}. 
Diverse techniques were used in obtaining input from human sources, including surveys~\cite{Merrill:2026Wearable}, interviews~\cite{Cui:2026Cgm}, and red-teaming~\cite{Arora:2025Healthbench}. However, few benchmarks combined different sources and techniques in diversifying their sources~\cite{Arora:2025Healthbench, Heydari:2025Anatomy}. 
Human input was often synthesized as parametrized templates that were then automatically populated to generate \taskeg{s}~\cite{Cardei:2025Dmbench, Merrill:2026Wearable, Tian:2026Lifeagentbench, Cui:2026Cgm, Christmann:2026Perqa}. In one notable case the input was used to simulate a full conversation~\cite{Arora:2025Healthbench}.
LLMs or human experts typically verified resulting \taskeg{s} against predefined criteria~(\eg logic and realism) or against the benchmark scope~(\eg domain relevance and answerability under the benchmark design). Only one benchmark verified \taskeg{s} by comparing them with queries collected from real users \cite{Christmann:2026Perqa}. Benchmarks often did not report how issues identified during verification were resolved or whether verification covered all \taskeg{s} or only a subset~\cite{Merrill:2026Wearable, Tian:2026Lifeagentbench, Cui:2026Cgm}. Even when \taskeg{s} were verified, issues such as logical inconsistencies persisted when verification did not include all relevant aspects. For example, \phia~\cite{Merrill:2026Wearable} reviewed \taskeg{s} for \qt{precision and comprehensibility} but did not report how these criteria were defined, the instructions reviewers followed, or how disqualified \taskeg{s} were handled. Our analysis of \task{s} revealed logical issues, such as misalignment between the operation of interest and the scope of operations (\eg \qt{How many days have I slept for 519 minutes or less today?}, `today' is meaningless here).

Task data were most commonly synthetically generated or drawn from existing datasets, and only two benchmarks collected data from real users~\cite{Cardei:2025Dmbench, Heydari:2025Anatomy}. Task data were also mainly passively sensed (\eg number of steps taken, CGM); only two benchmarks included self-reported data, such as meal and exercise logs~\cite{Cardei:2025Dmbench, Tian:2026Lifeagentbench}.

Benchmarks often did not report verification of task data~\cite{Cardei:2025Dmbench, Tian:2026Lifeagentbench, Cui:2026Cgm, Heydari:2025Anatomy, Healey:2024LLMcgm}. One notable quality control procedure appeared in \phia, where missing values were incorporated in synthetic wearable data to mirror patterns of missingness in real data~\cite{Merrill:2026Wearable}. Prior research has shown that missing data affects the data analysis performance of a \sut~\cite{Gu:2026Radar}. Such quality control can therefore lead to more realistic evaluation results (\ie it is more ecologically valid).

All benchmarks with reusable \evalsetting{s} were single-turn, and a \sut\ was often prompted with a \taskeg\ and its data. One slightly different response elicitation technique presented a conversation history ending with a user query as the \taskeg~\cite{Arora:2025Healthbench}.
Scoring methods were often defined by benchmark developers without reporting details of the creation process, and other sources of expertise, such as end users' lived experience or domain literature, were not used. Scoring criteria were often predefined as high-level descriptions, such as safety (\ie ``no risk of emotional, medical, or physical harm, including avoiding medical prognoses, or diagnoses'') and actionability~(\ie ``offering useful, practical guidance'')~\cite{Cardei:2025Dmbench}. Criteria were applied to responses by comparing them against these definitions~\cite{Cardei:2025Dmbench} or against rules that added points when a response met a desirable criterion and deducted points when it met an undesirable one~\cite{Arora:2025Healthbench}. The same criteria were often applied to all \taskeg{s} within a benchmark; in one exception, physicians wrote criteria specific to each \taskeg~\cite{Arora:2025Healthbench}. Only one benchmark verified its scoring method, examining the performance of its LLM grader and the relevance of criteria to \taskeg{s}~\cite{Arora:2025Healthbench}. Although that same benchmark did not report verification of other aspects of its scoring methods.

\subsection{Gaps in Benchmark Design for Personal Health Informatics}
\label{sec:results-gaps}
We find significant mismatches between the current landscape of benchmark design and the knowledge and guidelines of personal informatics literature. 
Existing benchmarks fail to consider many of the capabilities highlighted in empirical~(\eg \cite{Chopra:2025Genai}) or theoretical research on personal informatics~(\eg stages of tracking~\cite{Li:2010Stage, Epstein:2015Lived}, or key concepts around types of goals, tracking models, and data models~\cite{Sefidgar:2024Migrainetracker}). 
Moreover, their design choices are often based on assumptions that do not typically hold in individuals' lived experience.
They also fall short of incorporating documented aspects of the lived experience.
Taken together these shortcomings limit the extent to which benchmarks can successfully produce evidence on whether and how well LLMs support \phinfo\ activities.
Below we further unpack these observations.


\subsubsection{Benchmark Designs Do Not Align with Empirical Findings of Personal Health Informatics}
\label{sec:results-gaps-unaccounted}
Mismatches were present both in \task{s} benchmarks included and their \evalsetting{s}. 
Despite being commonly present in large numbers in benchmarks, \taskeg{s} under the \taskbehaviorvar\ category failed to cover the kinds of queries individuals commonly pose in \phinfo. For example, only one of the \nBenchmarks\ benchmarks included \taskeg{s} concerning variations in medication use which is especially of interest in chronic condition management as documented in past work~(\eg \cite{Sefidgar:2025Thesis}). 
\taskCorrelation\ \taskeg{s} also fell short of covering the diverse ways individuals frame their queries to understand potential relationships in their data. For example, individuals use different temporal alignments between measures, relating the occurrence of one measure to the presence of another concurrently, after a certain time~(\eg experiencing symptoms a day after alcohol consumption), or over a certain continued time~(\eg experiencing symptoms after two or more days of poor sleep)~\cite{Sefidgar:2025Thesis}. 
Beyond concurrent co-occurrence, benchmarks did not consider these other temporal alignments.

In addition to missing common user needs in \task\ design, both the content and language of \taskeg{s} were at times unrealistic. 
For example, queries used ``median'' or ``std'', words that are too technical for many end-users to use~(case
of median) or too distant from the language they actually use (\eg ``fluctuation'' rather than ``std''~\cite{Rey:2023Efficient}). 
Or, they included complex combinations of scoping~(\eg \qt{What was the total number of calories I burned during my last 5 runs within the past 30 days?}~\cite{Merrill:2026Wearable}) that very few end-users would use in a single query. To the contrary, evidence suggests individuals iteratively refine their queries, gradually adding additional complexity~\cite{Chopra:2025Genai}. 
Queries in each iteration are simple. For example, individuals 
explore their data by scoping it to a time window or an activity, or vaguely inquire without specifying a scope at all~\cite{Rey:2023Efficient}. 
Across these examples, there are unaccounted considerations for the design of \evalsetting{s}, including in criteria for checking the quality of \taskeg{s}~(\eg in using and operationalizing realism as a measure) or in the setup for response generation (\eg to allow multiple turns).
These observations surface a major concern: benchmarks did not collect evidence that mattered and they fell short at collecting evidence that did.

\subsubsection{Benchmarks Make Assumptions That Do Not Hold}
\label{sec:results-gaps-assume}
Benchmark \task{s} and \evalsetting{s} were often designed with the assumption that end-user queries were clearly worded and their personal data contained all the relevant information in a standard, clean, and clear representation that remained consistent over time. However, these assumptions are commonly untrue in real-world context, jeopardizing the ecological validity of the evidence benchmarks offer. 
\fatemeh{I'd suggest expanding the following paragraph based on the outline below:
\begin{itemize}
    \item Individuals struggle to articulate their needs.
    \item (Explain the natural interaction flow in more detail.) Empirical findings show that users shape their queries iteratively during conversations with LLMs by asking follow-up questions, adding more personal data, and refining their wording. (We can add an example from the Chopra paper: changing the query from losing weight to losing body fat, and adding step count data after multiple rounds.)
    \item In the real world, LLMs play a key role in what users take away from a conversation, not only by providing an answer but also by determining what is asked and what personal information is provided. \cite{Sayres:2026Context}
    \item Benchmarks do not test LLMs on this integral part of real-world interactions, clarifying queries and seeking additional context as needed, because they assume that models' responses to their \taskeg in a single turn are representative of what is involved in a real conversation. For example, an \taskpitoadvice query such as \qt{How do I improve my diet?}~\cite{Merrill:2026Wearable} can be answered in many ways depending on the user's needs and constraints, but this clarification capability cannot be tested, and the necessary information cannot be obtained, due to benchmark design limitations (e.g., single-turn settings).
\end{itemize}}
Contrary to the assumptions, many individuals struggle to articulate their needs~(\eg \cite{Schroeder:2018Migraine, Sefidgar:2025Thesis}), including in their interactions with LLMs~\cite{Chopra:2025Genai}. Their queries are often vague with the expectation that LLMs seek clarification, a capability benchmarks did not include or explicitly excluded~(\eg \cgmspeak\ labeled \taskeg{s} that require clarification as ``unanswerable''). The single-turn setting of the majority of benchmarks we reviewed made it impossible to consider this capability. 

Moreover, personal health data is heterogeneous. It includes multiple types of data with different underlying tracking models: some data may be recorded on a regular basis~(\eg daily), while others may be recorded on specific occasions~(\eg when symptoms occur). There is also evidence that tracking models change over time~\cite{Sefidgar:2025Thesis}. 
Heterogeneity of data and evolution of data generation processes have implications for data analysis~(\eg \cite{Sefidgar:2025Thesis}) 
that are neither part of \taskeg{s} for testing, nor in assessing responses of a \sut\ when scoring. For example, if there is a change in data generation process, systems should be tested on such tasks as detecting the change, and verifying it~(\eg by seeking user input). Existing benchmarks do not consider such testing. 

\subsubsection{Benchmarks Fail to Examine Key Capabilities for Personal Health Informatics}
\label{sec:results-gaps-capabilities}
A number of capabilities that benchmarks should demonstrate in LLM systems for personal health informatics given the field's literature were either absent or minimally covered by the existing benchmarks.
For example, very few or no \taskeg{s} existed to demonstrate if LLMs could help individuals prepare for tracking experience~(\eg in deciding what information is relevant to their goals), in configuring the experience~(\eg in translating concepts to tracking tool settings), 
or in lapsing from and resumption of tracking~(\eg in deciding `what' to retrieve from past tracking or `how' upon resumption). However, empirical studies document challenges in these activities and the desire that LLMs help address them~\cite{Chopra:2025Genai}.
Disambiguation of user queries or their personal data is an important capability across the stages~\cite{Sefidgar:2024Migrainetracker, Chopra:2025Genai}, including in data analysis, that was rarely considered by benchmarks. In the absence of evidence for this capability, it is unclear whether performance reports that benchmarks provide on data analysis would translate into real-world deployments. 
More broadly, benchmarks should examine capabilities for checking the assumptions 
and requirements around accomplishing a task~(\eg disambiguation of data, including its availability, for data analysis) and mitigating violations of assumptions and requirements~(\eg provision of guidance on changes to tracking to potentially make initially unavailable data available for analysis). To put it concretely, it is not enough for benchmarks to only examine if a \sut\ can answer ``how much sleep do I get on average?'' They have to also test that the system checks if sleep data is available to reliably answer the question, and if not, guides the user on how to get the data.
Moreover, stages of tracking are inter-related~\cite{Li:2010Stage} and individuals may be concurrently at different stages for their various needs~\cite{Sefidgar:2024Migrainetracker}. Current benchmarks' use of isolated queries within a single stage~(\ie data analysis) failed to surface whether and how well LLMs could account for the inter-relations in the tracking experience. For example, if they provide stage-specific support in a timely manner. 

\section{Discussion}
\label{sec: discussion}
We examined if benchmarks adequately establish LLM capabilities that support personal health informatics activities. 
To this end, we characterized the space of benchmark design choices and used this characterization as a language to map out and describe the landscape of existing benchmark design.
Comparing this landscape with the theoretical and empirical knowledge in personal health informatics, we identified gaps in how benchmarks incorporate domain understanding, their assumptions contrary to the lived experience of personal informatics, and the key capabilities they should examine in LLM systems but do not. Taken together, our observations demonstrate the limitations of existing benchmarks~(\eg lack of ecological validity) in showing LLMs support for personal health informatics activities.
Below we discuss the implications of our work for benchmark development~(\sect{sec:discussion-design}), opportunities it highlights for leveraging HCI and personal health informatics research~(\sect{sec:discussion-research}), and its limitations
~(\sect{sec:discussion-limits}).

\subsection{Characterization of Design Space As a Resource for Benchmark Development}
\label{sec:discussion-design}
The capabilities we noted as inadequately supported or completely missing in benchmarks based on personal health informatics knowledge are important to address in benchmark design. 
They are expected by end-users but LLM systems, especially the general-purpose ones, struggle to offer them effectively~\cite{Chopra:2025Genai}. 
It is thus important to improve benchmarking to help obtain reliable measurements of LLM systems' performance to guide progress. 
Facilitating this goal, our work offers resources for analysis, design, and documentation of benchmarks for personal health informatics.
Our characterization of the space of design choices in benchmarks can be used as a tool to analyze benchmarks for the kinds and limits of evidence they offer, similar to what we did for each of the benchmarks of our corpus.
Moreover, benchmark developers can use the design dimensions we identified to guide their thinking about decisions to make and resources to tap in making those decisions. 
Those same dimensions can also help in documenting the design of the benchmarks for transparency and reproducibility, as they provide clarity on the information to document and a structure for that documentation. 
Our work extends other efforts with the same objective~(\eg \cite{Liu:2024Ecbd}) by providing domain-specific understanding and guidance, adding to the growing body of domain-specific benchmark design~(\eg \cite{Li:2026Journalism, Palani:2026Lexara}). 


\subsection{Opportunities for Human-Centered and Domain-Informed Benchmark Design}
\label{sec:discussion-research}
Considering the current state and focus of benchmark design, we believe there needs to be a fundamental shift in the approach to benchmark design for personal health informatics. 
Given the diversity and complexity of this process, it is important to consider both \textit{what} activities people do and \textit{how} they do it. 
The latter is completely absent in existing designs but is exactly where personal informatics and HCI researchers are positioned to contribute. 
Both theories and empirical findings of personal informatics reflective of lived experience can be used to guide task design and generation as well as response generation procedures. 
For example, taking a goal-directed design approach~\cite{Schroeder:2019Goal, Sefidgar:2025Thesis}, benchmarks can aim to include \taskeg{s} that examine LLMs handling of data and goal evolution in collection, integration, and reflection on personal data. They can pair their task design with a multi-turn response generation process where different types of evolution are simulated~\cite{Sefidgar:2024Migrainetracker}.
More broadly, HCI methods can be leveraged to facilitate understanding and integration of lived experiences. 
For example, through such methods as scenario-based surveying in eliciting user input~\cite{Liao:2022Contextualized}. 
We note that domain-informed benchmark design we are advocating is complementary to other techniques~(\eg \cite{Hamna:2026Community}) that similarly aim to improve benchmarks. 


\subsection{Limitations}
\label{sec:discussion-limits}
Our examination of the state of benchmark design is by no means a comprehensive account of the shortcomings to address. It is primarily an opening for continued consideration of benchmark design in personal health informatics, and more broadly HCI. In fact, assessments of evaluation techniques, including benchmarks, are ongoing efforts. Benchmark evaluation and design should evolve as progress in theoretical and empirical understanding of the lived experience of personal health informatics surfaces new capabilities and considerations to incorporate. The language we have developed is an initial step in enabling these efforts.

We focused our investigation on personal health informatics activities and whether the relevant capabilities are part of benchmarking. There are many other inter-related dimensions to benchmark evaluation that we did not consider but are equally important. These include but are not limited to whether benchmarks adequately establish safety or proper handling of cultural differences. Future research should investigate these other dimensions.

\section{Conclusion}
\label{sec:conclusion}
Well designed benchmarks are powerful tools in driving LLM system development as they provide evidence on whether and how well systems support capabilities of interest.
We examined how well benchmarks are designed to assess LLM systems' ability to effectively and meaningfully support personal health informatics activities. Characterizing the design space of benchmark design, we identified shortcomings in incorporating domain knowledge in benchmarking of data analysis, the primary focus of existing benchmarks. 
We also documented problematic assumptions in benchmark design and highlighted that important capabilities, especially those pertaining to other stages of personal informatics, are currently missing in benchmarks.
We discussed the utility of our design space characterization as an analytic, decision-support, and documentation resource.
We also pointed out ways domain knowledge and HCI techniques can contribute to benchmark design efforts.







\bibliographystyle{ACM-Reference-Format}
\bibliography{references}

\FloatBarrier
\appendix
\section{Search Query}
\label{app:search-query}
\begin{figure}[H]
    \centering
    \includegraphics[width=0.8\linewidth]{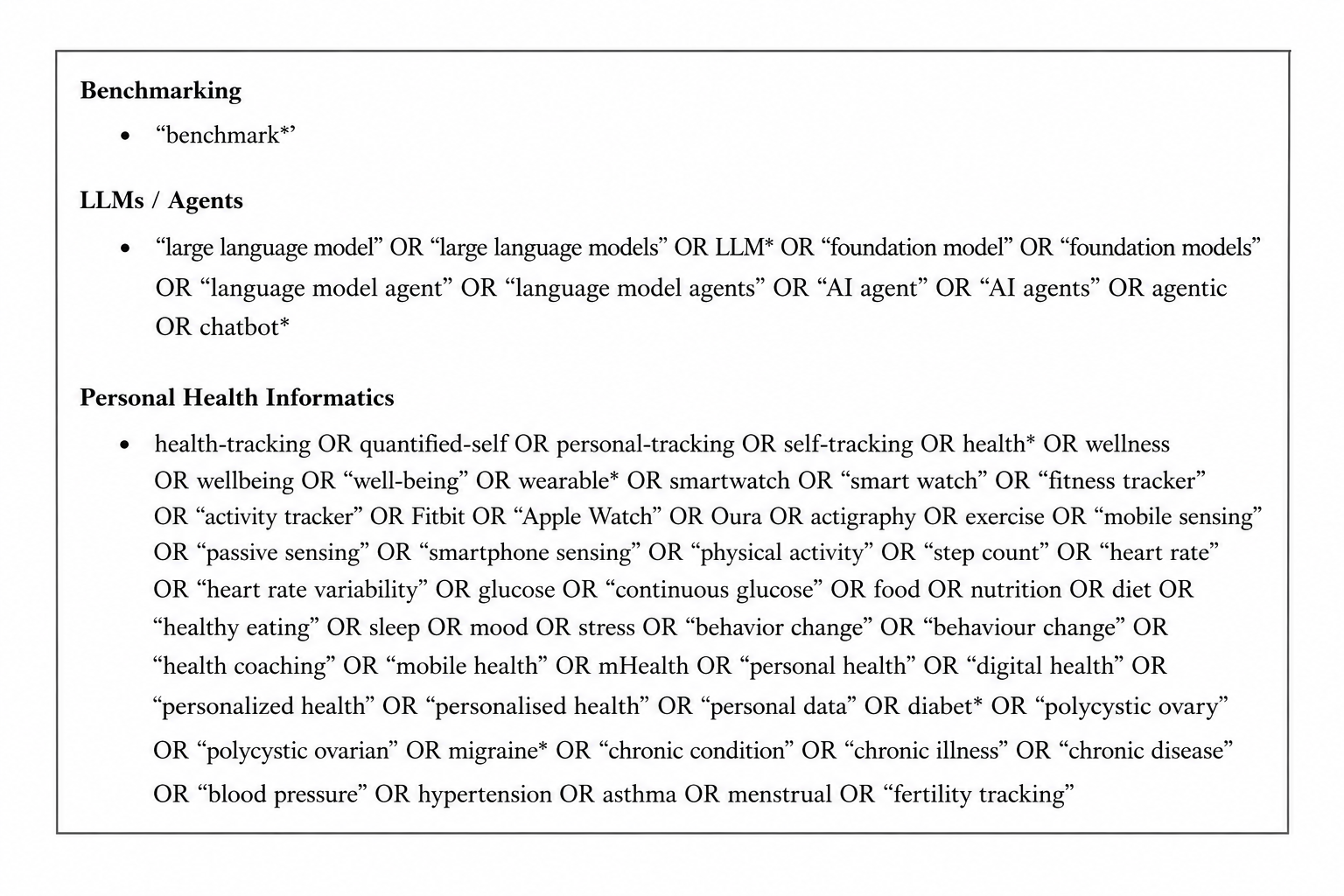}
    \caption{Database search query used to identify benchmarks. Together with papers identified through snowballing, 5,734 papers entered the screening process. We screened these papers against our inclusion and exclusion criteria (see \sect{sec:method-selection-criteria} and \sect{sec:method-selection-procedure}), manually reviewed 60 papers at full text, and arrived at the final corpus of \nBenchmarks\ benchmarks summarized in \tab{tab:benchmarks}.}
    \Description{
    Full database search query used to identify candidate publications for the review.
    The query combines groups of terms using Boolean operators. 
    The first group contains terms related to “benchmark* .
    The second group contains terms related to "large language model" OR "large language models" OR LLM* OR "foundation model" OR "foundation models" OR "language model agent" OR "language model agents" OR "AI agent" OR "AI agents" OR agentic OR chatbot* .
    The third group contains terms related to AND health-tracking OR quantified-self OR personal-tracking OR self-tracking
OR health* OR wellness OR wellbeing OR well-being  
OR wearable* OR smartwatch OR "smart watch" OR "fitness tracker" OR "activity tracker" OR Fitbit OR "Apple Watch" OR Oura OR actigraphy OR exercise OR "mobile sensing" OR "passive sensing" OR "smartphone sensing" OR "physical activity" OR "step count" OR "heart rate" OR "heart rate variability" OR glucose OR "continuous glucose" OR food OR nutrition OR diet OR "healthy eating" OR sleep OR mood OR stress OR "behavior change" OR "behaviour change" OR "health coaching" OR "mobile health" OR mHealth OR "personal health" OR "digital health" OR "personalized health" OR "personalised health" OR "personal data" OR diabet* OR "polycystic ovary" OR "polycystic ovarian" OR migraine* OR "chronic condition" OR "chronic illness" OR "chronic disease" OR "blood pressure" OR hypertension OR asthma OR menstrual OR "fertility tracking"”
    Terms within each concept group are connected with OR, while the concept groups are connected with AND.
    The complete query shown in the figure is:
    ``“benchmark* 
AND "large language model" OR "large language models" OR LLM* OR "foundation model" OR "foundation models" OR "language model agent" OR "language model agents" OR "AI agent" OR "AI agents" OR agentic OR chatbot* 
AND health-tracking OR quantified-self OR personal-tracking OR self-tracking
OR health* OR wellness OR wellbeing OR well-being  
OR wearable* OR smartwatch OR "smart watch" OR "fitness tracker" OR "activity tracker" OR Fitbit OR "Apple Watch" OR Oura OR actigraphy OR exercise OR "mobile sensing" OR "passive sensing" OR "smartphone sensing" OR "physical activity" OR "step count" OR "heart rate" OR "heart rate variability" OR glucose OR "continuous glucose" OR food OR nutrition OR diet OR "healthy eating" OR sleep OR mood OR stress OR "behavior change" OR "behaviour change" OR "health coaching" OR "mobile health" OR mHealth OR "personal health" OR "digital health" OR "personalized health" OR "personalised health" OR "personal data" OR diabet* OR "polycystic ovary" OR "polycystic ovarian" OR migraine* OR "chronic condition" OR "chronic illness" OR "chronic disease" OR "blood pressure" OR hypertension OR asthma OR menstrual OR "fertility tracking"”

.''
    }
    \label{fig:search-query}
\end{figure}

The database search query used in our review is shown in \fig{fig:search-query}.
\section{Prompt for Screening Papers}
\label{app:screening-prompt}

We used the following prompt to assist with screening papers against our inclusion and exclusion criteria:

\begin{tcolorbox}[
    breakable,
    colback=white,
    colframe=black,
    boxrule=0.5pt,
    arc=0pt,
    left=5pt,
    right=5pt,
    top=5pt,
    bottom=5pt,
    width=0.94\linewidth,
    center
]
\small

You are a systematic literature review assistant.

Your task is to screen papers for a review about LLM-based benchmarks on personal health data.

\textbf{INCLUSION criteria (ALL must be true):}

\begin{enumerate}
    \item The study evaluates LLMs directly or LLM-powered systems on tasks framed as an individual making an inquiry involving their self-tracked health data, which may be passively sensed (e.g., wearable data) or actively reported (e.g., self-logged records).

    \item The tasks involve natural-language interactions.

    \item The study offers material that can be reused to examine capabilities of LLMs or LLM-powered systems.
\end{enumerate}

\textbf{EXCLUSION criteria (ANY one is sufficient to exclude):}

\begin{enumerate}
    \item[\textbf{E1.}] The study focuses on clinically performed tasks such as disease diagnosis or treatment planning given clinical indications and input (e.g., adjusting medication dosage based on symptoms).

    \item[\textbf{E2.}] The study evaluates general personal health questions without a connection to personal data (e.g., learning about general guidelines for physical activity levels).

    \item[\textbf{E3.}] The tasks require only information such as demographics or general profiles (e.g., diet type or family history of health conditions), rather than longitudinal self-tracked data (e.g., food recommendations based on dietary restrictions and medical history).

    \item[\textbf{E4.}] The study primarily emphasizes sensing, i.e., the extraction of health markers, typically from raw physiological signals (e.g., deriving heart rate, heart-rate variability, or the presence of arrhythmia from ECG signals).
\end{enumerate}

\textbf{Examples of INCLUDED tasks:}

\begin{itemize}
    \item ``According to my Fitbit data, what was my average heart rate during running?''
    \item ``How is my sleep affected by my daily steps?''
    \item ``Based on my CGM data, what meal should I eat now?''
    \item ``Which activity burned more calories---yoga or swimming?''
    \item ``Tell me about anomalies in my active zone minutes last month.''
    \item ``How do my symptoms vary after starting a new treatment?''
\end{itemize}

\textbf{Examples of EXCLUDED tasks:}

\begin{itemize}
    \item ``Should I increase the dosage of my medication?'' (clinical task)
    \item ``Based on my symptoms, what is your diagnosis?'' (clinical task)
    \item ``What are side effects of [medication]?'' (general health question without personal data)
\end{itemize}

\textbf{Paper to screen:}

\textbf{Title:} \{title\}

\textbf{Abstract:} \{abstract\}

Respond \textbf{ONLY} with a JSON object---no markdown, no preamble---in exactly this format:

\texttt{\{"label": "<include|exclude|not\_sure>",}

\texttt{"reason": "<one concise sentence referencing the criteria>"\}}

\end{tcolorbox}

\section{Prompt for Classifying HealthBench Examples}
\label{app:healthbench-prompt}

We used the following prompt to identify HealthBench examples relevant to our review:

\begin{tcolorbox}[
    breakable,
    colback=white,
    colframe=black,
    boxrule=0.5pt,
    arc=0pt,
    left=5pt,
    right=5pt,
    top=5pt,
    bottom=5pt,
    width=0.94\linewidth,
    center
]
\small

You are classifying \taskeg{s} from the HealthBench benchmark for a review of LLM-based benchmarks involving personal health data.

For each HealthBench \taskeg, determine whether it should be included based on the following criteria.

\textbf{Include the example if:}

\begin{enumerate}
    \item The task is framed from the perspective of an individual making an inquiry involving their own self-tracked health data, or seeking support to set up or refine the collection of such data.

    \item The personal health data are longitudinal or repeated measurements that could be passively sensed or actively reported, such as wearable data, activity records, sleep records, glucose measurements, symptoms, bowel logs, diet logs, or other self-tracked records.
\end{enumerate}

\textbf{Exclude the example if:}

\begin{enumerate}
    \item The task is primarily clinician-facing, such as diagnosis, treatment planning, clinical documentation, or interpretation of clinical records for a healthcare professional.

    \item The task asks a general health question without requiring the individual's personal data.

    \item The task only provides static profile information such as age, sex, medical history, family history, or dietary preferences, rather than longitudinal self-tracked data.
\end{enumerate}

For each \taskeg, output:

\begin{itemize}
    \item \textbf{Classification:} Include or Exclude
\end{itemize}

\end{tcolorbox}

\section{Prompt for Mapping Benchmark \Taskeg{s} to Task Categories}
\label{app:task-mapping-prompt}

We used the following prompt to map benchmark \taskeg{s} to the task categories developed in our analysis:

\begin{tcolorbox}[
    breakable,
    colback=white,
    colframe=black,
    boxrule=0.5pt,
    arc=0pt,
    left=5pt,
    right=5pt,
    top=5pt,
    bottom=5pt,
    width=0.94\linewidth,
    center
]
\small

Your task is to classify benchmark \taskeg{s} using the categories below.

For each benchmark task example, assign the \textbf{single best-fitting category} based on the underlying information need and capability required to answer the task.

\begin{enumerate}

\item \textbf{Behavior Variations.}

Tasks that ask about values or patterns in an individual's personal health data. These tasks can generally be represented as an operation performed on a recorded measure within some scope of the data. Operations may involve retrieving values, summarizing records, or characterizing changes in the data.

\textit{Examples:} ``What's my longest sleep duration last month?''; ``What was my step count yesterday?''

\item \textbf{Correlation Analysis.}

Tasks that examine potential relationships between different aspects of an individual's personal health data, including relationships between health measures, time periods, or events.

\textit{Examples:} ``How does my sleep duration correlate with my daily steps?''; ``Could you explain why my steps fluctuate from weekdays to weekends?''

\item \textbf{Comparison.}

Tasks that consider differences between personal health data at two or more reference points. Comparisons may be made within an individual's own data or against an external reference, such as a population, demographic group, or guideline.

\textit{Example:} ``How does my exercise compare to pregnant women my age?''

\item \textbf{Data Quality Assessment.}

Tasks that require identifying limitations or problems in the available personal health data, such as missing, unreliable, or insufficient information.

\textit{Example:} ``Are there any artifacts in my data?''

\item \textbf{Advice Grounded in Personal Data.}

Tasks that assess combining information derived from an individual's personal health data with external knowledge to provide recommendations, suggested actions, or support for health-related decisions.

\textit{Example:} ``How do I improve my deep sleep?''

\item \textbf{Preparation and Setup.}

Tasks that involve support for personal data collection, such as creating tracking tools or deciding what and how to track.

\textit{Example:} ``Create a table to track my blood pressure at home, include guidance on understanding readings and when to contact healthcare provider.''

\end{enumerate}

\textbf{Instructions:}

Assign one category to each task. Base the assignment on what the user is asking the system to do with their personal health data.

\end{tcolorbox}

\section{Prompt for Mapping Behavior Variation Task Components}
\label{app:variation-mapping-prompt}

We used the following prompt to characterize the components of benchmark tasks categorized as Behavior Variations:

\begin{tcolorbox}[
    breakable,
    colback=white,
    colframe=black,
    boxrule=0.5pt,
    arc=0pt,
    left=6pt,
    right=6pt,
    top=6pt,
    bottom=6pt
]
\small

You are helping label benchmark \taskeg{} categorized as \textbf{Behavior Variations}. Behavior Variation tasks ask about values or patterns in an individual's personal health data and can be represented as \texttt{([operation], [measure], [scope])}, where \textit{operation} specifies what is done with the data, \textit{measure} specifies the recorded attribute being analyzed, and \textit{scope} specifies which records are included.

For each \taskeg{}, identify its \textbf{operation}, \textbf{measure}, and \textbf{scope}, then assign the corresponding variation label based on the observed values of \textbf{operation} and \textbf{scope}.

\textbf{Operation} is one of:

\begin{itemize}
    \item \textbf{aggregate}: performs an aggregation over a set of records, such as average, sum, count, minimum, or maximum.

    \item \textbf{value look up}: retrieves a specific value or set of values from one or more records.
\end{itemize}

\textbf{Measure} is the recorded value of an attribute of a behavior or state that the task asks about, such as sleep duration, step count, heart rate, workout speed, or blood glucose.

\textbf{Scope} specifies which records are included in the operation. Assign one of the following:

\begin{itemize}
    \item \textbf{global}: no scope filter is specified.

    \item \textbf{time window}: records are restricted to a date or time range.

    \textit{Example:} ``in the last 7 days''

    \item \textbf{threshold}: records are restricted according to a condition on a measure. The threshold may be numeric, categorical, or dynamically derived from the data.

    \textit{Examples:} ``on days I slept more than 7 hours''; ``on days when food\_meal\_labels = vegetables and fruits''; ``Tell me about anomalies in my active zone minutes last month.'' In the last example, the scope includes a dynamically derived threshold used to identify statistical outliers.

    \item \textbf{ordinal}: records are restricted based on order or ranking.

    \textit{Example:} ``my last five runs''

    \item \textbf{event-relative}: records are restricted based on their temporal relationship to an event or activity, such as occurring during, before, after, or since the event.

    \textit{Example:} ``my heart rate during the Euro Cup''

    \item \textbf{time window + ordinal}

    \textit{Example:} ``What was the duration of my most recent run within the last 7 days?''

    \item \textbf{time window + threshold}

    \textit{Example:} ``How many days in the past week did I sleep for 329 minutes or less?''
\end{itemize}

Assign exactly one of the following labels:

\begin{itemize}
    \item Variation ([aggregate], [measure]) globally
    \item Variation ([aggregate], [measure]) over time window
    \item Variation ([aggregate], [measure]) over threshold
    \item Variation ([aggregate], [measure]) over ordinal
    \item Variation ([aggregate], [measure]) over event-relative
    \item Variation ([aggregate], [measure]) over time window + ordinal
    \item Variation ([aggregate], [measure]) over time window + threshold
    \item Variation ([value look up], [measure]) globally
    \item Variation ([value look up], [measure]) over time window
    \item Variation ([value look up], [measure]) over threshold
    \item Variation ([value look up], [measure]) over ordinal
    \item Variation ([value look up], [measure]) over event-relative
    \item Variation ([value look up], [measure]) over time window + ordinal
    \item Variation ([value look up], [measure]) over time window + threshold
\end{itemize}

If a task cannot be mapped to any of the listed combinations of operation and scope, label it as \textbf{Couldn't map}.

\end{tcolorbox}

\FloatBarrier
\section{Details of \qt{\taskBehaviorVar}}
\label{app:variation-details}

\tab{tab:variation-components} provides the detailed structure and observed values used to characterize \qt{\taskBehaviorVar} \taskex{}.

\begin{table*}
\caption{Values observed in the components of \qt{\taskBehaviorVar} \taskex{}:
([operation], [measure], [scope]).}
\label{tab:variation-components}
\small
\renewcommand{\arraystretch}{1.15}

\begin{tabularx}{\textwidth}{
    @{}p{0.17\textwidth}
    p{0.18\textwidth}
    X@{}
}
\toprule
\textbf{Component} &
\textbf{Value} &
\textbf{Definition or observed values} \\
\midrule


\multirow{3}{*}{\textbf{Operation}}
& Value lookup &
Retrieves a value from one or multiple records within a scope. \\
\addlinespace[5pt]

& Aggregation &
Applies an aggregation over records within a scope, including
sum, average, minimum, maximum, count, median, dispersion
(\ie{} standard deviation and variance), mode, percentage,
and trend analysis. \\
\addlinespace[5pt]

\midrule


\multirow{6}{*}{\textbf{Measure}}
& Sleep &
Duration of overall sleep and sleep stages or states, including
deep sleep, REM sleep, light sleep, awake time, and in-bed time. \\
\addlinespace[5pt]

& Daily activity level &
Duration spent at different activity levels, including sedentary,
lightly active, moderately active, very active, fat burn, cardio,
and peak. \\
\addlinespace[5pt]

& Physical activity &
Duration and other measures associated with physical activity,
including date/time, elevation gain, distance, speed,
and steps. Observed activity types include walk, run, outdoor bike,
bike, aerobic workout, weights, elliptical, yoga, spinning,
treadmill, tennis, hike, pilates, workout, interval workout,
and swimming. \\
\addlinespace[5pt]

& Meals &
Meal category, subcategory, cooking style and carbohydrate intake. \\
\addlinespace[5pt]

& Heart activity &
Resting heart rate, heart-rate variability. \\
\addlinespace[5pt]

& Glycemic activity &
Blood glucose
(CGM mg/dL) and associated measures. \\
\addlinespace[5pt]

& Stress &
Stress management score. \\

& Medication use &
Insulin intake. \\

\midrule


\multirow{6}{*}{\textbf{Scope}}
& Global &
Uses all available records without applying a filter,
\eg{} \qt{What was my mean blood glucose?} \\
\addlinespace[5pt]

& Time window &
Restricts records to a specified date or time interval,
\eg{} \qt{What unique activities have I done in the past 7 days?} \\
\addlinespace[5pt]

& Threshold &
Restricts records based on meeting a condition. Observed threshold
types include:
\textbf{numeric}, \eg{} \qt{How many days have I slept less than 8 hours?};
\textbf{categorical}, \eg{} \qt{How many days did I consume food in the
Vegetables and Fruits category?}; and
\textbf{dynamic}, where the threshold is derived from the data itself,
such as a statistical boundary. \\
\addlinespace[5pt]

& Ordinal &
Restricts records to those defined by a ranking or ordering,
\eg{} \qt{What was my average elevation gain for my last four aerobic workouts?} \\
\addlinespace[5pt]

& Event-relative &
Restricts records based on their temporal relationship to an event
or activity,
\eg{} \qt{What's the total distance I cycled since I started my last job?} \\
\addlinespace[5pt]

& Combined &
Combines multiple scope types. Observed combinations included
time window + threshold and time window + ordinal,
\eg{} \qt{How many days during the past 21 days did I sleep for less than
374 minutes?} and
\qt{What was the duration of my most recent run within the last 30 days?} \\

\bottomrule
\end{tabularx}

\Description{
A table describing the observed values for the three components of behavior metric variation task examples: operation, measure, and scope.
For operation, value lookup retrieves a value from one or multiple records within a scope. Aggregation applies an aggregation over records within a scope, including sum, average, minimum, maximum, count, median, dispersion such as standard deviation and variance, mode, percentage, and trend analysis.
For measure, sleep includes duration of overall sleep and sleep stages or states, including deep sleep, REM sleep, light sleep, awake time, and in-bed time. Daily activity level includes duration spent at different activity levels, including sedentary, lightly active, moderately active, very active, fat burn, cardio, and peak. Physical activity includes duration and other measures such as date and time, elevation gain, distance, calories, speed, and steps, across activities including walking, running, outdoor biking, biking, aerobic workouts, weights, elliptical, yoga, spinning, treadmill, tennis, hiking, pilates, workouts, interval workouts, and swimming. Meals include meal category, subcategory, and cooking style. Heart and glycemic activity include resting heart rate, heart-rate variability, and blood glucose measured in CGM milligrams per deciliter. Stress includes stress management score.
For scope, global uses all available records without applying a filter. Time window restricts records to a specified date or time interval. Threshold restricts records based on a numeric, categorical, or dynamically derived condition. Ordinal restricts records based on a ranking or ordering. Event-relative restricts records according to their temporal relationship to an event or activity. Combined scope applies multiple scope types together, with observed combinations including time window plus threshold and time window plus ordinal.
}

\end{table*}

\FloatBarrier









\section{Details of Benchmark \EvalSetting{s}}
\label{app:settings-details}

The following tables report the detailed values observed for the dimensions and sub-dimensions of benchmark \evalsetting{s}. 
\tab{tab:task-creation-verification} reports observed values for Task Creation and Task Verification; 
\tab{tab:task-data-settings} reports observed values for Task Data Collection/Generation and Task Data Verification and Quality Control; and 
\tab{tab:response-generation-scoring} reports observed values for Response Generation and Response Scoring.

\begin{table}[H]
\centering
\caption{Observed values for Task Creation and Task Verification.}
\label{tab:task-creation-verification}
\small
\renewcommand{\arraystretch}{1.25}
\setlength{\extrarowheight}{1.5pt}

\begin{tabularx}{\columnwidth}{@{}p{0.23\columnwidth} X@{}}
\toprule
\textbf{Sub-dimension} & \textbf{Observed values} \\
\midrule

\multicolumn{2}{@{}l}{\textbf{Task Creation}} \\[2pt]

Source of information (expertise)
& Model evaluation experts (benchmark developers) // physicians: clinical expertise // personal health research experts: topic expertise // end users: lived experience // existing datasets: literature support. \\
\arrayrulecolor{rowgray}
\cmidrule(lr){1-2}
\arrayrulecolor{black}

Information collected
& Templates // interaction scenarios // ``as-is'' user queries // models' vulnerabilities. \\
\arrayrulecolor{rowgray}
\cmidrule(lr){1-2}
\arrayrulecolor{black}

Process of information collection
& Surveys // interviews // log analysis: logs of web searches, AI chatbots, and online forums // red-teaming of models // expert authoring. \\
\arrayrulecolor{rowgray}
\cmidrule(lr){1-2}
\arrayrulecolor{black}

Task creator
& Model evaluation experts (benchmark developers) // LLMs. \\
\arrayrulecolor{rowgray}
\cmidrule(lr){1-2}
\arrayrulecolor{black}

Transformation of information into tasks
& Direct use without changes // processing: populating templates by assigning values to parameters; role-playing multi-turn conversations using the collected information. \\

\midrule

\multicolumn{2}{@{}l}{\textbf{Task Verification}} \\[2pt]

Verifier
& LLMs // human experts: technical and health experts; model evaluation experts (benchmark developers). \\
\arrayrulecolor{rowgray}
\cmidrule(lr){1-2}
\arrayrulecolor{black}

Aspect verified
& Linguistic quality, logic, and soundness // distinctness from other tasks within the benchmark // similarity to human-generated queries // alignment with benchmark scope: domain relevance; answerability given benchmark design; task--data grounding // precision // comprehensibility. \\
\arrayrulecolor{rowgray}
\cmidrule(lr){1-2}
\arrayrulecolor{black}

Verification procedure
& Assessment against predefined criteria // comparison with human-generated tasks. \\
\arrayrulecolor{rowgray}
\cmidrule(lr){1-2}
\arrayrulecolor{black}

Remediation
& Iterative refinement // removal of disqualified tasks. \\
\arrayrulecolor{rowgray}
\cmidrule(lr){1-2}
\arrayrulecolor{black}

Scope of verification
& All tasks // subset of tasks. \\

\bottomrule
\end{tabularx}
\Description{
A table summarizing observed values for two benchmark design dimensions: Task Creation and Task Verification.
For Task Creation, the source of information includes model evaluation experts or benchmark developers, physicians contributing clinical expertise, personal health research experts contributing topic expertise, end users contributing lived experience, and existing datasets providing literature support.
The information collected includes templates, interaction scenarios, user queries used as-is, and model vulnerabilities.
Information collection processes include surveys, interviews, analysis of logs from web searches, AI chatbots, and online forums, model red-teaming, and expert authoring.
Tasks are created by model evaluation experts or benchmark developers and by large language models.
Collected information is transformed into tasks either through direct use without changes or through processing, including populating templates by assigning values to parameters and role-playing multi-turn conversations using the collected information.
For Task Verification, verifiers include large language models and human experts, including technical experts, health experts, and model evaluation experts or benchmark developers.
Verified aspects include linguistic quality, logic and soundness, distinctness from other tasks within the benchmark, similarity to human-generated queries, alignment with benchmark scope including domain relevance, answerability given benchmark design and task-data grounding, precision, and comprehensibility.
Verification procedures include assessment against predefined criteria and comparison with human-generated tasks.
Remediation includes iterative refinement and removal of disqualified tasks.
Verification may be applied to all tasks or to a subset of tasks.
}
\end{table}

\begin{table}[H]
\centering
\caption{Observed values for Task Data Collection/Generation and Task Data Verification and Quality Control.}
\label{tab:task-data-settings}
\small
\renewcommand{\arraystretch}{1.25}
\setlength{\extrarowheight}{1.5pt}

\begin{tabularx}{\columnwidth}{@{}p{0.25\columnwidth} X@{}}
\toprule
\textbf{Sub-dimension} & \textbf{Observed values} \\
\midrule

\multicolumn{2}{@{}l}{\textbf{Task Data Collection/Generation}} \\[2pt]

Collection/generation process
& Collection from: existing datasets (potentially with adaptations such as reformatting); people for the purpose of evaluation // synthetic generation using: neural-network models; statistical models; physiological simulator. \\
\arrayrulecolor{rowgray}
\cmidrule(lr){1-2}
\arrayrulecolor{black}

Data recording type
& Passively sensed // actively self-reported. \\
\arrayrulecolor{rowgray}
\cmidrule(lr){1-2}
\arrayrulecolor{black}

Data source
& Real // synthetic. \\
\arrayrulecolor{rowgray}
\cmidrule(lr){1-2}
\arrayrulecolor{black}

Data domain
& CGM // heart rate // physical activity // sleep // stress // meals. \\
\arrayrulecolor{rowgray}
\cmidrule(lr){1-2}
\arrayrulecolor{black}

Input data scope
& One day // within a week // within a month // from a few weeks to a few months // multi-month // multi-year. \\
\arrayrulecolor{rowgray}
\cmidrule(lr){1-2}
\arrayrulecolor{black}


Data format
& Structured: tabular (CSV, SQL tables); JSON // unstructured: text data embedded in the prompt. \\

\midrule

\multicolumn{2}{@{}l}{\textbf{Task Data Verification and Quality Control}} \\[2pt]

Verification criteria
& Realism // response generation feasibility. \\
\arrayrulecolor{rowgray}
\cmidrule(lr){1-2}
\arrayrulecolor{black}

Quality-control procedures
& Benchmark developers checking against an existing persona // missing-value injection: patterns of missing data observed in real records are incorporated into simulated wearable data to mirror sporadic gaps // temporal aggregation: data are converted into coarser representations such as morning/afternoon/evening segments or 30-minute intervals. \\

\bottomrule
\end{tabularx}
\Description{
A table summarizing observed values for Task Data Collection/Generation and Task Data Verification and Quality Control.
For Task Data Collection/Generation, data are collected from existing datasets, potentially with adaptations such as reformatting, or from people specifically for evaluation, and may also be synthetically generated using neural-network or statistical models.
Data recording types include passively sensed and actively self-reported data.
Data sources may be real or synthetic.
Observed data domains include continuous glucose monitoring, heart rate, physical activity, sleep, stress, and meals.
Input data scopes range from one day, within a week, within a month, and from a few weeks to a few months, to multi-month and multi-year periods.
Data formats include structured representations such as tabular CSV files, SQL tables, and JSON, as well as unstructured text data embedded in prompts.
For Task Data Verification and Quality Control, verification criteria include realism and response generation feasibility.
Quality-control procedures include benchmark developers checking data against an existing persona, injecting missing values based on missing-data patterns observed in real records to reproduce sporadic gaps in simulated wearable data, and temporally aggregating data into coarser representations such as morning, afternoon, and evening segments or 30-minute intervals.
}
\end{table}

\begin{table}[H]
\centering
\caption{Observed values for Response Generation and Response Scoring. \phia{}, \cgmspeak{}, \pha{}, and \llmcgm{} are excluded, as their scoring methods are tied to the systems they were built for and cannot be applied to other systems.}
\label{tab:response-generation-scoring}
\small
\renewcommand{\arraystretch}{1.25}
\setlength{\extrarowheight}{1.5pt}

\begin{tabularx}{\columnwidth}{@{}p{0.25\columnwidth} X@{}}
\toprule
\textbf{Sub-dimension} & \textbf{Observed values} \\
\midrule

\multicolumn{2}{@{}l}{\textbf{Response Generation}} \\[2pt]

Conversation length
& Single-turn. \\
\arrayrulecolor{rowgray}
\cmidrule(lr){1-2}
\arrayrulecolor{black}

Model input
& A single query + its data // an existing conversation as \ataskex{}: potentially embedding the personal data. \\
\arrayrulecolor{rowgray}
\cmidrule(lr){1-2}
\arrayrulecolor{black}

Response elicitation
& Prompting the \sut{} with a single task example and its data to generate one response // providing an existing conversation context: appending the final user query and eliciting the \sut{}'s response. \\
\arrayrulecolor{rowgray}
\cmidrule(lr){1-2}
\arrayrulecolor{black}

Part of response scored
& Final generated response. \\
\arrayrulecolor{rowgray}
\cmidrule(lr){1-2}
\arrayrulecolor{black}

Response format
& Numeric answer // short-text response: e.g., yes/no // free-form text response. \\

\midrule

\multicolumn{2}{@{}l}{\textbf{Response Scoring}} \\[2pt]

Scoring method creator
& Model evaluation experts (benchmark developers) // physicians. \\
\arrayrulecolor{rowgray}
\cmidrule(lr){1-2}
\arrayrulecolor{black}

Grader
& LLM grader // evaluation script. \\
\arrayrulecolor{rowgray}
\cmidrule(lr){1-2}
\arrayrulecolor{black}

Scoring criteria
& Accuracy // groundedness // safety // clarity // actionability // completeness // context awareness // communication quality // instruction following. \\
\arrayrulecolor{rowgray}
\cmidrule(lr){1-2}
\arrayrulecolor{black}

Criteria operationalization
& Formula-based definition: comparison against ground truth; application of rules with points assigned based on whether the rule applies // free-form criterion definition. \\
\arrayrulecolor{rowgray}
\cmidrule(lr){1-2}
\arrayrulecolor{black}

Scoring scope
& Task-specific: criteria or criterion definitions are customized to individual tasks or task categories // benchmark-wide: fixed criteria and definitions are applied across tasks. \\
\arrayrulecolor{rowgray}
\cmidrule(lr){1-2}
\arrayrulecolor{black}

Metrics
& Mean criterion-point score // pass rate // accuracy. \\
\arrayrulecolor{rowgray}
\cmidrule(lr){1-2}
\arrayrulecolor{black}

\multirow{3}{=}{Scoring verification}
& \textbf{What is verified:} grader reliability // relevance of rubric criteria. \\
\arrayrulecolor{rowgray}
\cmidrule(l{6pt}r{6pt}){2-2}
\arrayrulecolor{black}

& \textbf{Verifier:} human expert. \\
\arrayrulecolor{rowgray}
\cmidrule(l{6pt}r{6pt}){2-2}
\arrayrulecolor{black}

& \textbf{Verification procedure:} grader reliability: human experts independently grade model responses, and agreement between the LLM grader and human raters is compared with human--human agreement // rubric relevance: multiple physicians independently judge whether a criterion applies to a task, and the criterion is retained when a majority agree. \\

\bottomrule
\end{tabularx}
\Description{
A table summarizing observed values for Response Generation and Response Scoring. PHIA, CGM Speak, PHA, and LLM-CGM are excluded because their scoring methods are tied to the systems they were developed for and cannot be applied to other systems.
For Response Generation, observed conversations are single-turn.
Model input consists either of a single query together with its data or an existing conversation used as a task example, which may embed personal data.
Responses are elicited either by prompting the system under test with a single task example and its data to generate one response or by providing an existing conversation context, appending the final user query, and eliciting the system's response.
Only the final generated response is scored.
Response formats include numeric answers, short-text responses such as yes or no, and free-form text responses.
For Response Scoring, scoring methods are created by model evaluation experts or benchmark developers and by physicians.
Graders include LLM graders and evaluation scripts.
Observed scoring criteria include accuracy, groundedness, safety, clarity, actionability, completeness, context awareness, communication quality, and instruction following.
Criteria are operationalized either through formula-based definitions, including comparison against ground truth or application of rules with assigned points, or through free-form criterion definitions.
Scoring scope may be task-specific, where criteria or their definitions are customized to individual tasks or task categories, or benchmark-wide, where fixed criteria and definitions are applied across tasks.
Observed metrics include mean criterion-point score, pass rate, and accuracy.
Scoring verification examines grader reliability and the relevance of rubric criteria.
Physicians serve as verifiers.
For grader reliability, human experts independently grade model responses, and agreement between the LLM grader and human raters is compared with human--human agreement.
For rubric relevance, multiple physicians independently judge whether a criterion applies to a task, and the criterion is retained when a majority agree.
}
\end{table}

\FloatBarrier

\section{Characterization of \cgmspeak}
\label{app:char-cgmspeak}
\begin{figure*}[p]
  \centering
  \includegraphics[width=\textwidth,height=0.88\textheight,keepaspectratio]{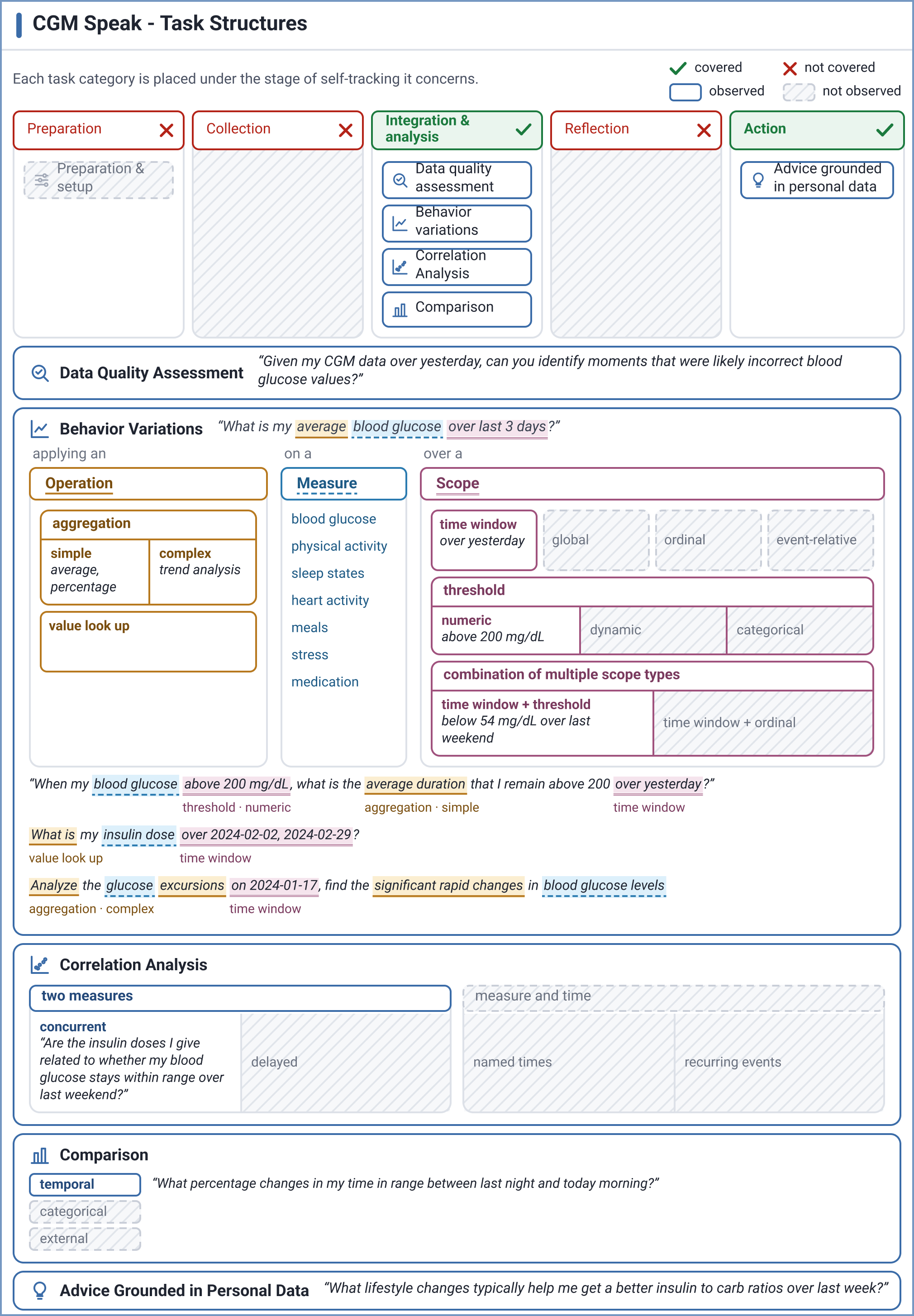}
\caption{Task categories observed in \cgmspeak, covering the integration and action stages of self-tracking~\cite{Li:2010Stage}; no tasks were observed for the preparation, collection, or reflection stages. Each category is shown with the components and subtypes described in \sect{sec:result-design-space}, along with examples. In \taskBehaviorVar, both operation types, aggregation and value look-up, are observed, while among scope types only time window and numeric threshold appear. In \taskCorrelation, only the measure--measure subtype is observed (\eg insulin dose and time in range), with no temporal alignment specified between the two measures, and measure--time correlations are absent. In \taskComparison, only temporal reference points are observed (\eg last night vs.\ this morning).}
  \Description{Card showing the task categories covered by CGM-Speak, each with one example task.}
  \label{fig:cgmspeak-tasks}
\end{figure*}

\begin{figure*}[p]
  \centering
  \includegraphics[width=\textwidth,height=0.88\textheight,keepaspectratio]{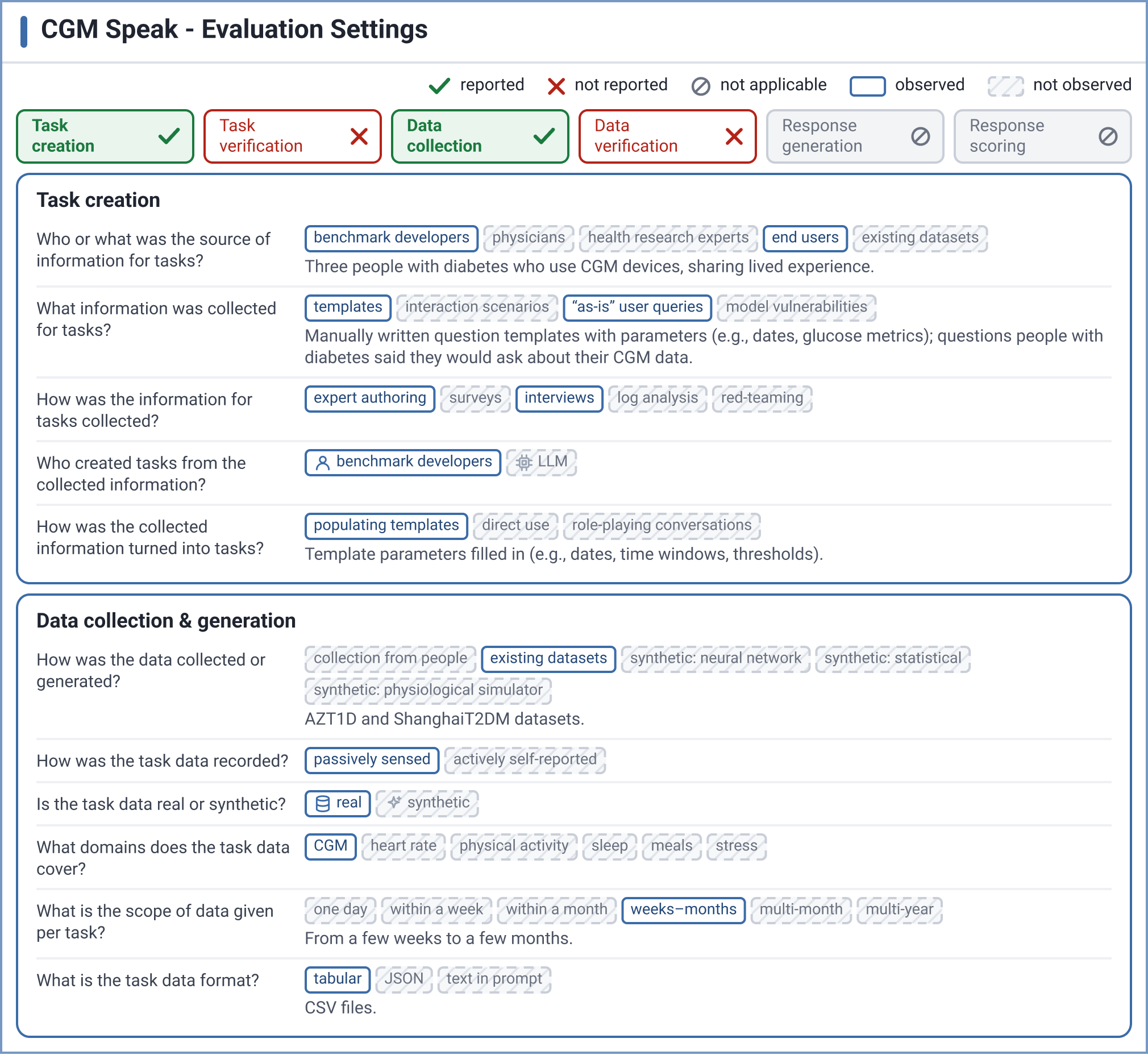}
  \caption{Evaluation settings of \cgmspeak, with task creation and task data collection reported; task verification and task data verification are not reported, and response generation and response scoring are not applicable because \cgmspeak's scoring method is not reusable for evaluation of other systems. Each dimension is shown with the sub-dimensions described in \sect{sec:result-design-space}, which characterize the design decisions made in building the benchmark, along with the observed values. In task creation, both benchmark developers and end users (three people with diabetes who use CGM devices) are sources of information, and templates and as-is user queries collected through expert authoring and interviews are turned into tasks by populating templates. In task data collection, real, passively sensed CGM data spanning a few weeks to a few months are drawn from existing datasets and provided in tabular format.}
  \Description{Card showing the evaluation settings of CGM-Speak, with observed and not-observed values for task creation and task data collection, and status markers for the remaining dimensions.}
  \label{fig:cgmspeak-settings}
\end{figure*}

\cgmspeak's task categories and evaluation settings are shown in \fig{fig:cgmspeak-tasks} and \fig{fig:cgmspeak-settings}, respectively. Characterization of other benchmarks in the corpus are provided in the supplementary material. 

\FloatBarrier


\end{document}
\endinput